\documentclass[%
 reprint,
superscriptaddress,
 amsmath,amssymb,
 aps,footinbib,
]{revtex4-1}

\usepackage[colorlinks=true,linkcolor=blue,urlcolor=blue,citecolor=blue]{hyperref}

\usepackage{overpic}
\usepackage{upgreek}
\usepackage{graphicx}
\usepackage{dcolumn}
\usepackage{bm}

\usepackage{xcolor}

\definecolor{mhs}{rgb}{0,0,0}
\definecolor{john}{rgb}{0,0,0}
\definecolor{johnnew}{rgb}{0,0,0}
\definecolor{mhsnew}{rgb}{0,0,0}
\definecolor{newall}{rgb}{0,0,0}
\definecolor{allnew}{rgb}{0,0,0}

\begin{document}

\title{Ferrimagnetic Oscillator Magnetometer}

\author{John F. Barry}
\email{john.barry@ll.mit.edu}
\affiliation{MIT Lincoln Laboratory, Lexington, Massachusetts 02421, USA}
\author{Reed A. Irion}
\affiliation{MIT Lincoln Laboratory, Lexington, Massachusetts 02421, USA}
\author{Matthew H. Steinecker}
\affiliation{MIT Lincoln Laboratory, Lexington, Massachusetts 02421, USA}
\author{Daniel K. Freeman}
\affiliation{MIT Lincoln Laboratory, Lexington, Massachusetts 02421, USA}
\author{Jessica J. Kedziora}
\affiliation{MIT Lincoln Laboratory, Lexington, Massachusetts 02421, USA}
\author{Reginald G. Wilcox}
\affiliation{MIT Lincoln Laboratory, Lexington, Massachusetts 02421, USA}
\affiliation{Massachusetts Institute of Technology, Cambridge, Massachusetts 02139, USA}
\author{Danielle A. Braje}
\affiliation{MIT Lincoln Laboratory, Lexington, Massachusetts 02421, USA}

\date{\today}

\begin{abstract}
Quantum sensors offer unparalleled precision, accuracy, and sensitivity for a variety of measurement applications. We report a compact magnetometer based on a ferrimagnetic sensing element in an oscillator architecture that circumvents challenges common to other quantum sensing approaches such as limited dynamic range, limited bandwidth, and dependence on vacuum, cryogenic, or laser components. The device exhibits a fixed, calibration-free response governed by the electron gyromagnetic ratio. Exchange narrowing in the ferrimagnetic material produces sub-MHz transition linewidths despite the high unpaired spin density ($\sim 10^{22}$ cm$^{-3}$). The magnetometer achieves a minimum sensitivity \textcolor{mhsnew}{of 100}~fT/$\sqrt{\text{Hz}}$ to AC magnetic fields of unknown phase and a sensitivity below 200~fT/$\sqrt{\text{Hz}}$ over a bandwidth $\gtrsim \! 1$~MHz. By encoding magnetic field in frequency rather than amplitude, the device provides a dynamic range in excess of 1~mT. The passive, thermal initialization of the sensor's quantum state requires only a magnetic bias field, greatly reducing power requirements compared to laser-initialized quantum sensors. With additional development, this device promises to be a leading candidate for high-performance magnetometry outside the laboratory, and the oscillator architecture is expected to provide advantages across a wide range of sensing platforms.
\end{abstract}

\maketitle
\onecolumngrid
\urlstyle{same}
For the published version, refer to Physical Review Applied, DOI: \href{https://doi.org/10.1103/PhysRevApplied.19.044044}{doi.org/10.1103/PhysRevApplied.19.044044}
\\
\twocolumngrid

\section{Introduction}
In recent years, tremendous experimental effort has advanced quantum sensors~\cite{degen2017quantumsensing} \textcolor{mhsnew}{using} unpaired electron spins embedded in solid-state crystals. These solid-state sensors employ electron paramagnetic resonance to achieve measurement precision and accuracy comparable to their atomic counterparts, but with advantages such as smaller sensing volumes, compatibility with a wide range of ambient conditions, and fixed sensing axes provided by a rigid crystal lattice. The most-developed solid-state quantum sensing platform uses negatively-charged nitrogen-vacancy (NV) centers in diamond as sensitive magnetic field probes~\cite{taylor2008high,degen2008scanning}. Such sensors have been used to detect or image biological targets~\cite{fescenko2019diamond,lesage2013optical,davis2018mapping,barry2016optical,webb2021detection,arai2022millimetrescale}, single proteins~\cite{lovchinsky2016nuclear,shi2015single}, NMR species~\cite{kehayias2017solution,glenn2018high,bucher2020hyperpolarization,smits2019two,aslam2017nuclear}, 
individual spins~\cite{sushkov2014magnetic,rugar2015proton,pelliccione2014two,sushkov2014alloptical}, and condensed matter phenomena~\cite{jenkins2020imaging,casola2018probing,bertelli2020magnetic,lenz2021imaging,broadway2020imaging}. 

Though recent efforts have focused on optically-active paramagnetic defects~\cite{chatzidrosos2017miniature,acosta2010broadband,eisenach2021cavity,barry2020sensitivity,hopper2018spin}, ferrimagnetic materials offer distinct advantages for quantum sensors. Ferrimagnetic materials provide higher unpaired electron spin densities than their solid-state paramagnetic counterparts~\cite{kimball2016precessing}, for example $\sim 10^{22}$ cm$^{-3}$ versus $\sim 10^{16}-10^{19}$ cm$^{-3}$, while the strong coupling of the exchange interaction mitigates the dipolar resonance broadening observed in high-defect-density paramagnetic materials~\cite{bauch2018ultralong,bauch2020decoherence}. Importantly, initialization of ferrimagnetic spins into the desired quantum state requires only a bias magnetic field, without the need for active optical initialization.

Consequently, magnetic sensors employing spin-wave interferometry in ferrimagnetic films~\cite{balynsky2017magnetometer, balinskiy2020spinwave} or ferrimagnetic resonance (FMR) in spheres~\cite{hempstead1964precise,carpenter1982phase,beaumont2019ferrimagnetic} or films~\cite{inoue2011investigating,kaya2015yig,koda2019highly,wen2022ferromagnetic} have been investigated, including demonstrations with pT/$\sqrt{\text{Hz}}$-level sensitivity. Using ferrimagnetic materials, classical sensors such as fluxgates~\cite{koshev2021evolution,vetoshko2003epitaxial,vetoshko2016fluxgate} and Faraday-rotation-based devices~\cite{doriath1982sensitive,holthaus2006magnetic} have achieved sensitivities down to $40$ fT/$\sqrt{\text{Hz}}$  and 10 pT/$\sqrt{\text{Hz}}$, respectively. Additionally, ferrimagnetic materials have long found commercial use in tunable microwave filters~\cite{carter1961magnetically,degrasse1959lowloss} and oscillators~\cite{chang1967yigtuned,omori1969octave,ollivier1972microwave}. Despite these well-developed commercial technologies however, magnetometry schemes for ferrimagnetic materials have not previously employed a self-sustaining oscillator architecture to encode magnetic fields in the output waveform frequency rather than amplitude \footnote{\textcolor{newall}{We note that Ref. [35] incorporates positive feedback in an oscillator configuration into their magnetometer device, but the positive feedback is only used to enhance the change in observed microwave power depending on the value of magnetic field. The device is not designed to produce sustained oscillations as magnetic fields are varied, nor to measure fields by monitoring the frequency of the circulating microwave power.}}. We find this architecture provides crucial advantages in performance, capabilities, and simplicity of a magnetometer device. 

Here we report a magnetometer using FMR as the magnetically-sensitive frequency discriminator in an electronic oscillator. With this construction, the frequency of the output voltage signal tracks the FMR frequency, which varies linearly with the applied magnetic field. This ferrimagnetic oscillator magnetometer exhibits a minimum sensitivity of 100~fT/$\sqrt{\text{Hz}}$ to magnetic fields near 100 kHz and sensitivities below 200~fT$/\sqrt{\text{Hz}}$ \textcolor{mhsnew}{from} 3 kHz to 1 MHz\textcolor{mhsnew}{.} As the device encodes the measured magnetic field directly in frequency, superior dynamic range is achieved relative to devices employing amplitude encoding. In addition, the sensor head is simple, compact, and lower power than existing quantum magnetometers of comparable sensitivity.  
\vfill

\section{Oscillator Architecture}\label{sec:oscillatorarchitecture}

Quantum sensors based on atomic vapors or electron spins in solid-state crystals operate by localizing resonances which vary with a physical quantity of interest. For example, the ambient magnetic field may be determined by measuring a ferrimagnetic material's uniform precession frequency~\cite{kitte1948theory}, the paramagnetic resonance frequency of NVs in diamond~\cite{oort1988optically}, or the hyperfine resonance frequency of an alkali vapor~\cite{dupontroc1969detection}. Several experimental techniques have been developed for this task, from \textcolor{mhsnew}{continuous-wave} absorption~\cite{acosta2010broadband} or dispersion~\cite{eisenach2021cavity,crescini2020cavity} measurements to pulsed protocols such as Ramsey~\cite{Ramsey1950molecular} or pulsed ESR~\cite{dreau2011avoiding} schemes. In all these methods, externally-generated electromagnetic fields manipulate the spin system, and the resonance location is determined from the system's resulting response. 

As an alternative to probing the spin system with external signals, however, an oscillator architecture can be arranged to generate a microwave (MW) signal that directly encodes the spin resonance location. Such an oscillator consists of two main components: a frequency discriminator and a gain element, arranged in a \textcolor{mhs}{feedback} loop.

The frequency discriminator can be constructed by coupling input MW signals to the discriminator's output through the quantum spins. If the discriminator's input and output are each coupled to the quantum spin resonance, but not directly to each other, the resulting frequency discriminator will pass frequencies near the spin resonance $\omega_y$ while rejecting all others.

The needed gain can be provided by an ordinary RF amplifier; by amplifying the frequency discriminator's output and returning a fraction of this signal to the discriminator's input, sustained self-oscillation can be realized~\cite{rubiola2009phasebook}. Because only frequencies near the spin resonance $\omega_y$ are transmitted through the frequency discriminator, the resultant oscillation frequency $\omega_c$ closely tracks the spin resonance. 

Thus, the oscillator architecture eliminates the need for an external RF source. The limited component count of the oscillator architecture is advantageous for compactness and design simplicity. In addition, the oscillator architecture encodes the spin resonance in frequency, which can offer greater dynamic range and \textcolor{mhs}{improved} linearity compared to amplitude-encoded measurements~\cite{limes2020portable}; dynamic range is particularly important for a magnetometer, where, for example, detection of a 100 fT signal in Earth's $\sim 0.1$ mT field requires a dynamic range $\sim 10^9$.

\textcolor{john}{For an oscillator \textcolor{mhs}{to operate} at steady state, losses through the frequency discriminator and other elements \textcolor{mhs}{must be exactly} compensated by the amplifier, \textcolor{mhs}{producing} unity gain around the oscillator loop. Additionally, the phase-length around the oscillator loop must equal an integer number of wavelengths at the steady-state oscillator output frequency. Together, these requirements constitute the Barkhausen criterion, and with reasonable assumptions the requirements result in Leeson's equation \cite{robins1984phase,everard2012simplified,rubiola2009phasebook,rhea2010discrete}, an empirical model of phase noise amplitude spectral density applicable to a wide range of oscillators. Leeson's equation is given by}
\begin{equation}\label{eqn:leesonphasenoisevoltage}
\mathcal{L}^\frac{1}{2}(f_m) = \sqrt{\frac{1}{2}\left[\frac{f_L^2}{f_m^2}+1\right] \left[ \frac{f_c}{f_m} +1 \right] \left[\frac{F k_B T}{P_s}\right]},
\end{equation}
where $\mathcal{L}^\frac{1}{2}(f_m)$ is the single-sideband phase noise amplitude spectral density at offset frequency $f_m$ from the carrier, $f_L$ is the Leeson frequency (equal to the frequency discriminator's loaded half width at half maximum linewidth), $f_c$ is \textcolor{mhsnew}{the} $1/f$ flicker noise corner~\cite{rubiola2009phasebook,boudot2012phase,robins1984phase}, $P_s$ is the input power to the sustaining amplifier, $T$ is the temperature, $k_B$ is Boltzmann's constant, and $F$ is the \textcolor{john}{oscillator's measured} wideband noise factor. \textcolor{mhs}{Roughly, Leeson's equation expresses the phase noise created by amplified white thermal noise (the final bracketed factor), enhanced within the bandwidth of the frequency discriminator via regeneration (the first bracketed factor), and further enhanced by flicker noise below the noise corner of the amplifier (the second bracketed factor). Additional details of oscillator phase noise are discussed in Supplemental Material (SM) Sec.~B \footnote{See Supplemental Material \label{SMRef} at [link to be inserted] for \textcolor{mhsnew}{additional} details of \textcolor{mhsnew}{the oscillator phase noise, YIG properties, crystal anisotropy, S-parameters for YIG transmission filter, oscillator magnetometer theory of operation, magnetometer noise, intrinsic linewidth measurement, magnetization tip angle, demodulation (magnetic field recovery), and test field calibration}}} In \textcolor{mhs}{Sec.}~\ref{sec:ferrimagnetic} we show the magnetometer's noise floor is proportional to $f_m \times \mathcal{L}^\frac{1}{2}(f_m)$, establishing the oscillator's phase noise as the principal determinant of magnetometer sensitivity.

\begin{figure*}[t]

\begin{overpic}[height=3.25in]{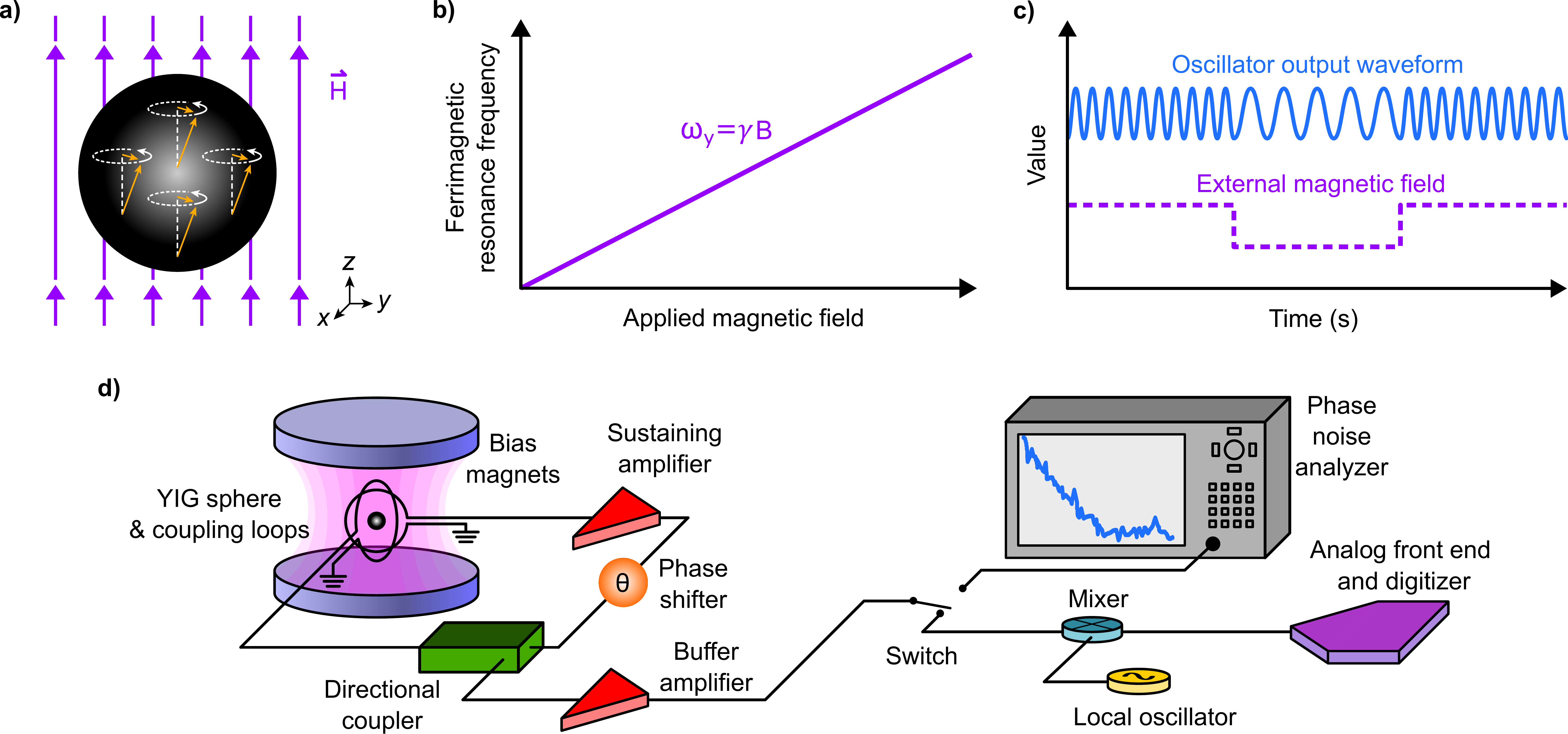}

\end{overpic}

\caption{\textbf{Oscillator magnetometer principles of operation.} (a) In the presence of a uniform external magnetic field, the spins of a ferrimagnetic sphere precess in phase. (b)  The resonance frequency of the uniform precession mode varies linearly \textcolor{mhsnew}{with applied} magnetic field. (c) By using the ferrimagnetic resonance as a frequency discriminator, an oscillator can be constructed where the oscillation frequency tracks the ferrimagnetic resonance frequency. (d) Experimental schematic as described in the main text.}\label{fig:deviceschematic}
\end{figure*}

\section{Ferrimagnetic Resonance}\label{sec:ferrimagnetic}

The material with the narrowest known \textcolor{john}{ferrimagnetic} resonance linewidth and lowest known spin-wave damping is yttrium iron garnet (YIG), a synthetic, insulating crystal ferrimagnet with chemical composition Y$_3$Fe$_5$O$_{12}$. Other attractive aspects of YIG are low acoustic damping, less than that of quartz, and well-developed growth processes which yield samples of high crystal quality~\cite{kolokolov1984magnon}. Consequently, YIG is the prototypical material for cavity spintronics research, and is used in magnon-cavity coupling experiments~\cite{tabuchi2014hybridizing,goryachev2014high,zhang2014strongly,bourhill2016ultrahigh,flower2019experimental,tabuchi2014hybridizing,awschalom2021quantum}, magneto-acoustic coupling studies~\cite{zhang2016cavity,li2018magnon}, hybrid quantum circuits~\cite{wolski2020dissipation,quirione2017resolving}, and axion searches~\cite{crescini2020cavity}.

In crystallographically-perfect YIG, five of every twenty lattice sites, equivalent to one unit formula Y$_3$Fe$_5$O$_{12}$, are populated by trivalent iron (Fe$^{3+}$, electronic spin $S=5/2$). The five trivalent iron atoms occupy three tetrahedral lattice sites and two octahedral lattice sites. Strong superexchange interactions, mediated by oxygen ions between the iron ions, align the three tetrahedral Fe$^{3+}$ antiparallel to the two \textcolor{mhsnew}{octahedral} Fe$^{3+}$ in the absence of thermal excitation. The strong coupling between nearby electronic spins results in collective spin behavior, including resonances between collective spin states which are observed as ferrimagnetic resonances. The strong spin-spin coupling also results in exchange-narrowing of the ferrimagnetic resonances,  allowing sub-MHz transition linewidths despite the high unpaired spin density \mbox{$\sim 10^{22}$ cm$^{-3}$}. Narrower resonances are desired to achieve better oscillator phase noise performance. Additional relevant properties of YIG are detailed in \textcolor{mhs}{SM Sec.~C \cite{Note2}}.

Kittel's formula for the uniform precession frequency of ferromagnetic resonance~\cite{kitte1948theory} is \mbox{$\omega_y = \sqrt{[\gamma B_z\!+\!(N_y\!-\!N_z)\gamma \mu_0 M_z][\gamma B_z\!+\!(N_x\!-\!N_z)\gamma \mu_0 M_z]}$},
where $\gamma$ is the electron gyromagnetic ratio\textcolor{mhs}{;} $\mathbf{B} = B_z\hat{z}$ is the applied magnetic field and defines the system's $\hat{z}$ axis\textcolor{mhs}{;} \textcolor{john}{$M_z$ is the magnetization along $\hat{z}$, where $M_z$ is assumed equal to the saturation magnetization $M_s$ with no MW power applied}\textcolor{mhs}{;} $N_x$, $N_y$, and $N_z$ are the demagnetization factors\textcolor{mhs}{;} and all quantities are in SI units. \textcolor{mhs}{Demagnetization factors characterize the shape-dependent reduction in internal \textcolor{john}{magnetic} field due to the magnetization~\cite{osborn1945demagnetizing}.} For a spherical sample, $N_x \!=\! N_y \!=\! N_z =\frac{1}{3}$, and Kittel's formula becomes
\begin{equation}\label{eqn:sphereresonance}
\omega_y = \gamma B_z.
\end{equation}
Kittel's above formula neglects crystal anisotropy, but such effects can be treated perturbatively if needed, as detailed in \textcolor{mhs}{SM Sec.~L \cite{Note2}}. 

A ferrimagnetic resonance can be used to implement the frequency discriminator discussed in \textcolor{mhs}{Sec.}~\ref{sec:oscillatorarchitecture}\textcolor{mhs}{, as shown in Fig.~~\ref{fig:deviceschematic}}. Consider two orthogonal circular coupling loops with a small ferrimagnetic sphere centered at the intersection of the \textcolor{john}{coupling} loop axes, as shown in Fig.~\ref{fig:deviceschematic}d. In the presence of an externally applied DC magnetic bias field $\mathbf{B} = B_0\hat{z}$, the magnetic domains within the sample align along $\hat{z}$, producing a net magnetization. A MW drive signal with angular frequency $\omega_d\approx\omega_y$, applied to the input coupling loop, causes the sphere's magnetization to precess about the $\hat{z}$ axis~\cite{lax1962microwave}\textcolor{mhs}{, as shown in Fig.~~\ref{fig:deviceschematic}a}. This precessing magnetization then inductively couples to the output coupling loop, and the transmission scattering parameter $S_{21}$ obeys 
\begin{equation}\label{eqn:s21yig}
S_{21} = \frac{\sqrt{\kappa_{1}\kappa_{2}}}{i(\omega_d-\omega_y) +\frac{\kappa_{0}+\kappa_{1}+\kappa_{2}}{2}}e^{-i\frac{\pi}{2}},
\end{equation}
where $\kappa_{0}$, $\kappa_{1}$, and $\kappa_{2}$ are the unloaded FMR linewidth, input coupling rate, and output coupling rate, respectively, in angular frequency units (see \textcolor{mhs}{SM Sec.~D \cite{Note2}}), and the $\pi/2$ phase retardation arises from the gyrator action of the ferrimagnetic material. The power transmission $|S_{21}|^2$ exhibits a Lorentzian line shape, with a maximum at the FMR frequency $\omega_y$ and a loaded full-width half-maximum (FWHM) linewidth  $\kappa_L \equiv \kappa_0+\kappa_1+\kappa_2$.

As discussed above, changes in the external DC magnetic field alter the FMR frequency $\omega_y$ and therefore the oscillator output frequency. The FMR frequency also responds to AC magnetic fields; the mechanism by which AC fields alter the magnetometer output waveform is discussed in \textcolor{mhs}{SM Sec.~F \cite{Note2}}. Operationally, AC magnetic fields are encoded as frequency modulation of the oscillator's output waveform. For example, an AC magnetic field with root-mean-square (rms) amplitude $B_\text{sen}^\text{rms}$ and angular frequency $\omega_m$ produces sidebands at $\pm \omega_m$ relative to the oscillator carrier frequency when applied parallel to $\mathbf{B}_0$. These two sidebands each exhibit a carrier-normalized amplitude of
\begin{equation}\label{eqn:response}
s = \frac{\gamma B_\text{sen}^\text{rms}}{\sqrt{2}\omega_m}. 
\end{equation}
The oscillator magnetometer's sensitivity can then be determined from the sideband amplitude and the measured phase noise  $\mathcal{L}^\frac{1}{2}(f_m)$, which represents the background against which the sidebands are discerned; see \textcolor{mhs}{SM Sec.~F~and~G~\cite{Note2}}. The expected sensitivity is
\begin{equation}\label{eqn:sensitivitynoleeson}
\eta(f_m) = \frac{\sqrt{2}f_m}{\gamma/(2\pi)} \mathcal{L}^\frac{1}{2}(f_m). 
\end{equation}
We note a striking feature of the oscillator magnetometer architecture: assuming the oscillator phase noise is well-described by Leeson's equation (Eqn.~\ref{eqn:leesonphasenoisevoltage}), the signal $s \propto 1/\omega_m = 1/(2\pi f_m)$ and the phase noise amplitude spectral density $\mathcal{L}^\frac{1}{2}(f_m)$ are expected to exhibit nearly identical scaling within a range of frequencies between \textcolor{mhsnew}{the noise} corner $f_c$ and the Leeson frequency $f_L$. Thus, the sensitivity of the device versus frequency $f_m$ is expected to be approximately flat for $f_c \lesssim f_m \lesssim f_L$, \textcolor{mhsnew}{as discussed} in \textcolor{mhs}{SM Sec.~G \cite{Note2}}.

\section{Experimental Setup}

\textcolor{john}{While all presently commercially-available YIG oscillators~\cite{microlambda2022electromagnetic,teledyne2022tiny} employ a reflection architecture, the device here employs a transmission (feedback) architecture~\cite{korber19933p67,sweet2006novel,sweet2006miniature,sweet2014wide,delden2019low}. The transmission oscillator} is constructed from four components connected in a serial \textcolor{john}{feedback} loop as shown in \textcolor{mhs}{Fig.~\ref{fig:deviceschematic}d}: the FMR frequency discriminator which only passes signals near the ferrimagnetic resonance~$\omega_y$, a directional coupler to sample the oscillator waveform for device output, a sustaining amplifier to provide the needed gain, and a mechanical phase shifter to ensure the Barkhausen criterion is satisfied~\cite{rubiola2009phasebook}. 

The device's sensing element is a 1-mm-diameter YIG sphere mounted on the end of an insulating ceramic rod. As shown in Fig.~\ref{fig:deviceschematic}, two circular coupling loops in the $xz$ and $yz$ planes inductively couple input and output MW signals to the YIG sphere. The coupling loops are mounted orthogonal to each other so that $S_{21}$ transmission occurs only at the FMR frequency and is suppressed elsewhere.  The values of $\kappa_0$, $\kappa_1$, and $\kappa_2$ are determined by simultaneously measuring the S-parameters $S_{11}$ and $S_{21}$ of the FMR frequency discriminator using a vector network analyzer\textcolor{mhsnew}{; see} \textcolor{mhs}{SM Sec.~E \cite{Note2}}.  We find $\kappa_0 = 2\pi \times 790$ kHz, $\kappa_1 \!=\! 2\pi \times 315$ kHz, and $\kappa_2\! =\! 2\pi\times\! 405$ kHz. The total loaded linewidth is then $\kappa_L \!\equiv\! \kappa_0\!+\!\kappa_1\!+\!\kappa_2 = 2\pi \times\! 1.510$ MHz, corresponding to a loaded quality factor $Q_L \!=\! \frac{\omega_y}{\kappa_L} \!=\! 3300$ and a predicted Leeson frequency of $f_L = \frac{1}{2}\frac{\kappa_L}{2\pi} = 755$ kHz.

Two cylindrical permanent magnets positioned symmetrically relative to the YIG sphere create a  uniform bias magnetic field $\mathbf{B_0} = B_0\hat{z}$ of approximately $0.178$~T, as \textcolor{mhs}{depicted} in Fig.~\ref{fig:deviceschematic}. \textcolor{john}{This value of $B_0$ is more than sufficient to saturate the sphere's magnetization}, so that the response is governed by $\omega_y(t) = \gamma B(t)$. The YIG sphere is aligned along the zero temperature compensation axis, as discussed in \textcolor{mhs}{SM Sec.~L \cite{Note2}}. With this bias magnetic field, the oscillation frequency is $\approx 2\pi \!\times \!5$ GHz.

The YIG sphere's precessing magnetization continuously induces a sinusoidal voltage on the output coupling loop at the magnetization's precession frequency. This MW voltage signal is first amplified and then mechanically phase shifted before passing through a 6 dB directional coupler, as shown in Fig.~\ref{fig:deviceschematic}d. The directional coupler's through port directs the MW signal back to the input coupling loop, inductively coupling the MW signal back to the YIG's precessing magnetization and closing the oscillator \textcolor{john}{feedback} loop. The mechanical phase shifter is adjusted to minimize the device phase noise, which is measured in real time.

Under operating conditions, the input power to the sustaining amplifier is $P_s\approx 3$ dBm. The sustaining amplifier has a measured gain of 10 dB at $P_s = 3$ dBm so that, after accounting for $\approx\!2$ dB of additional loss, $\approx 11$~dBm of MW power is delivered to the input coupling loop. \textcolor{john}{This MW power is estimated to tip the magnetization by $\approx$ 0.1 radians from the z axis; see \textcolor{mhs}{SM Sec.~H \cite{Note2}}.}

The signal sampled by the 6 dB directional coupler is first amplified by a buffer amplifier and then sent to either a phase noise analyzer for diagnostics and device optimization, or to a mixer which downconverts the signal to an intermediate frequency, $\omega_i$, in the MHz range appropriate for a digitizer. The downconverted signal is demodulated to recover the magnetic field time series as described in \textcolor{mhs}{SM Sec.~I \cite{Note2}}. \textcolor{mhs}{All experiments are performed with the device in an unshielded laboratory environment.}

\begin{figure*}[t]
\hspace{-3mm}
\begin{minipage}[b]{0.32\textwidth}
\begin{overpic}[height=1.79in]{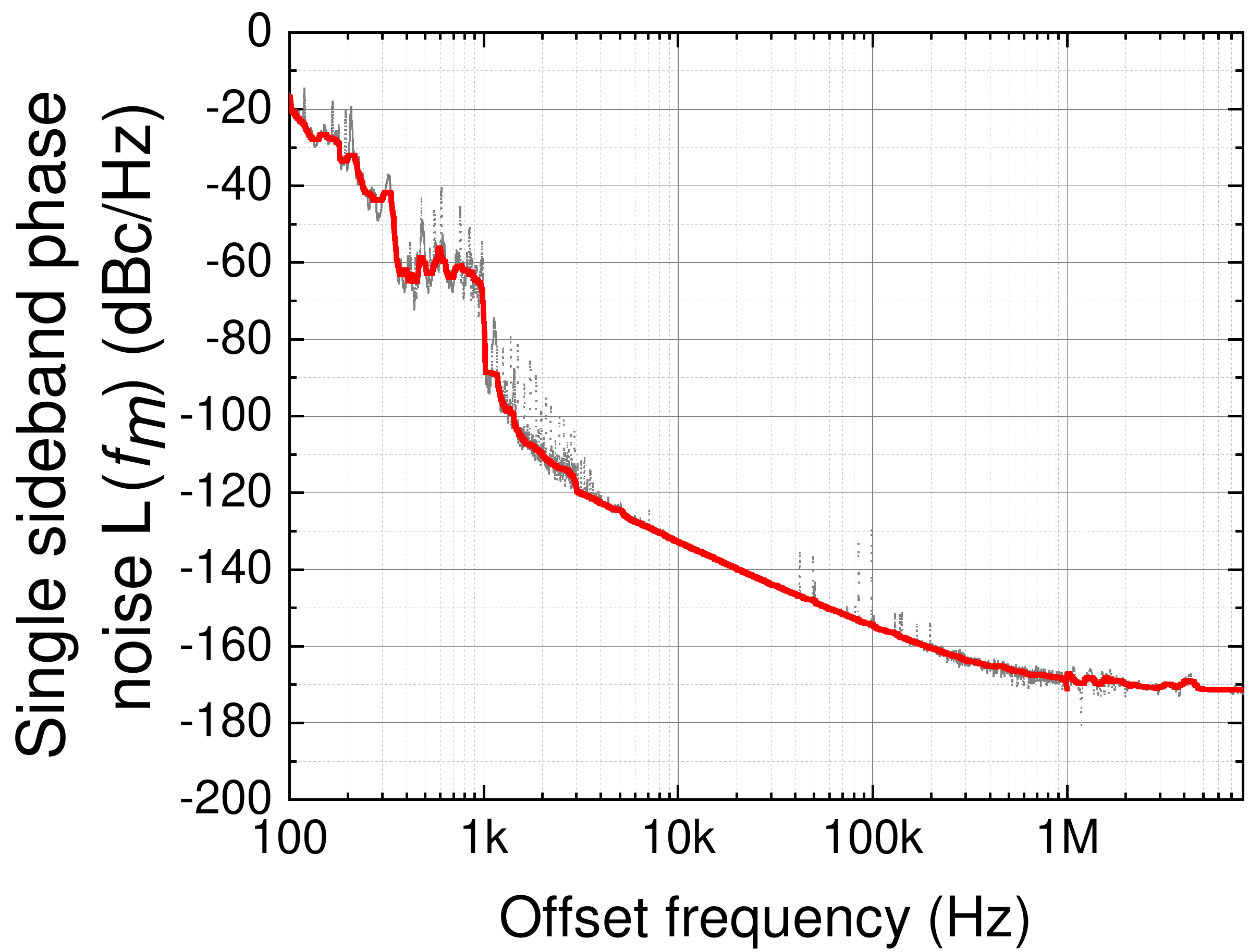} 
\put(8,78){\textbf{a)}}
\end{overpic}
\end{minipage}
\;\;
\begin{minipage}[b]{0.32\textwidth}
\begin{overpic}[height=1.79in]{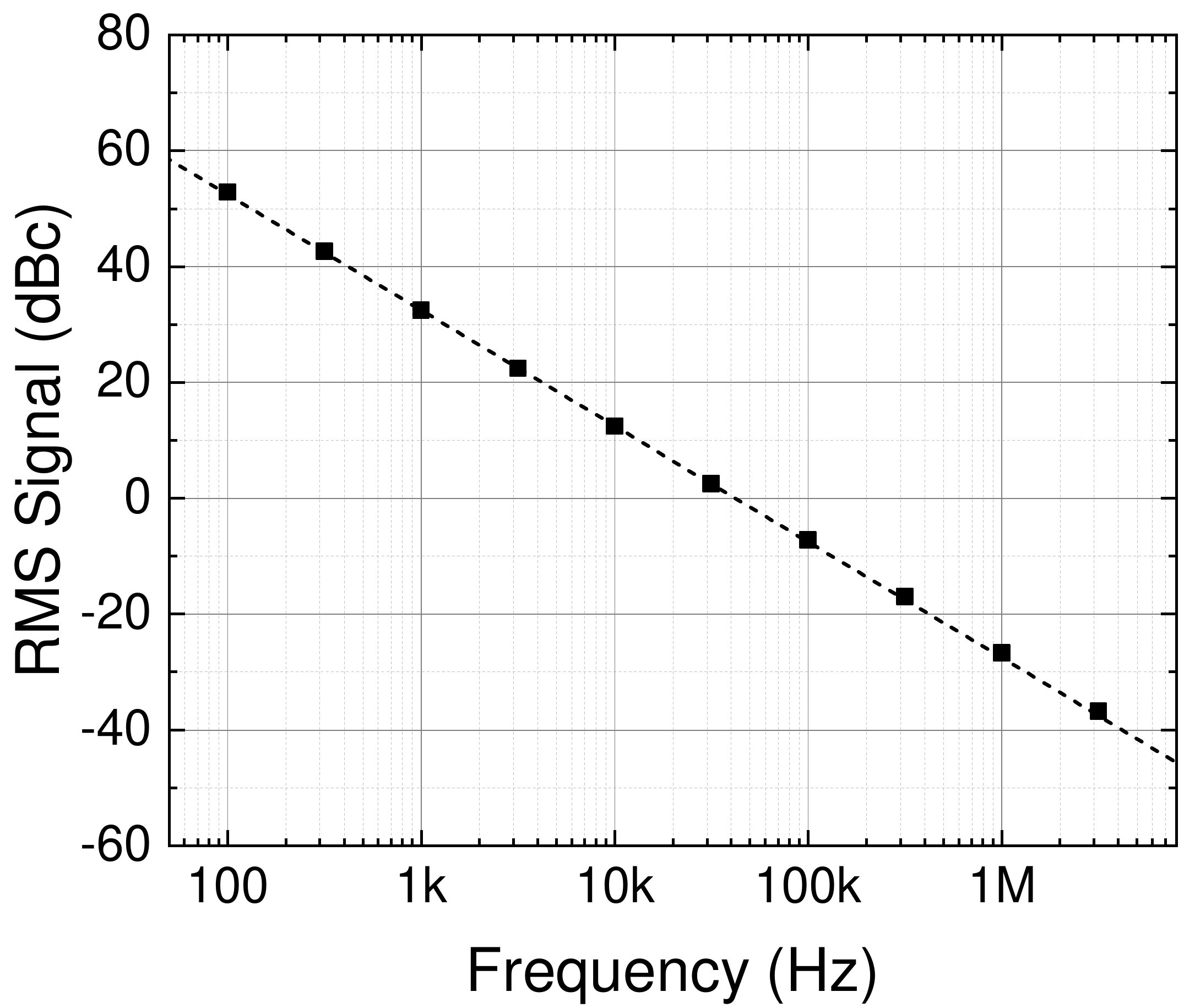} 
\put(-3,88){\textbf{b)}}
\end{overpic}
\end{minipage}
\begin{minipage}[b]{0.32\textwidth}
\begin{overpic}[height=1.82in]{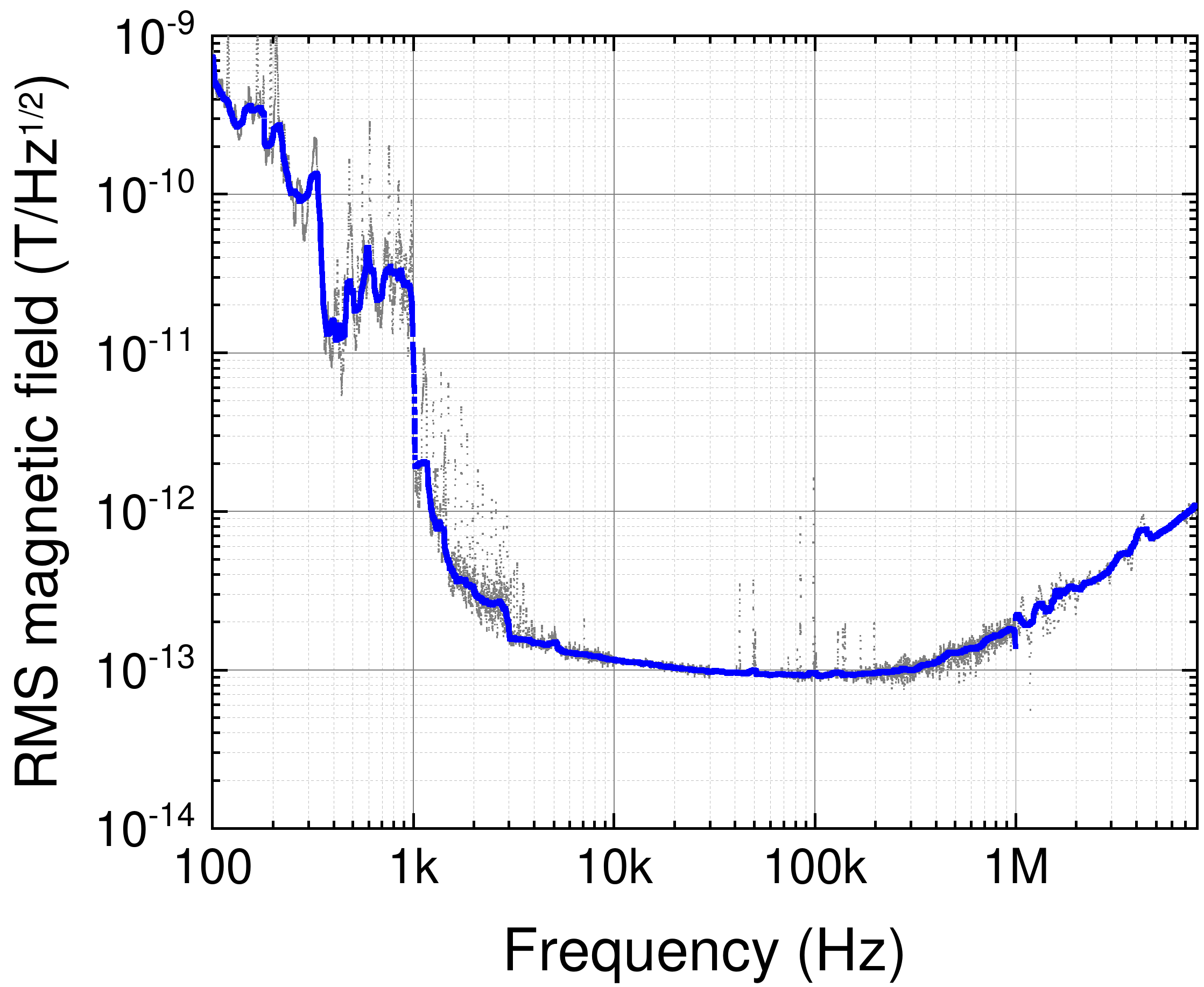} 
\put(1,83){\textbf{c)}}
\end{overpic}
\end{minipage}
\caption{\textbf{Performance of the ferrimagnetic oscillator magnetometer.} (a) Single-sideband phase noise power spectral density $\mathcal{L}(f_m)$ of the ferrimagnetic oscillator magnetometer. The single-sideband phase noise is $-132.8$ dBc/Hz and $-154.4$~dBc/Hz at 10 kHz and 100 kHz offsets from the carrier, respectively. Red depicts smoothed data while the raw phase noise data is gray. (b) Measured response ($\blacksquare$) to a 2.12 $\upmu$T rms AC magnetic field applied along $\hat{z}$, in agreement with that predicted by Eqn.~\ref{eqn:response} (\rule[.8mm]{1mm}{.3mm}\;\rule[.8mm]{1mm}{.3mm}\;\rule[.8mm]{1mm}{.3mm}). (c) Single-sided \textcolor{mhs}{magnetic field} amplitude spectral density of the ferrimagnetic oscillator magnetometer. The device achieves a minimum sensitivity of approximately 100~fT$/\sqrt{\text{Hz}}$ at frequencies near 100 kHz, with sensitivities below 200 fT$/\sqrt{\text{Hz}}$ \textcolor{mhsnew}{from} 3 kHz to 1 MHz. Blue depicts smoothed data while the raw data is gray. \textcolor{john}{\textcolor{mhs}{We note that by convention} the \textcolor{mhs}{single-sideband} quantity $\mathcal{L}(f)$ \textcolor{mhs}{is the positive-frequency} half of the double-\textcolor{mhs}{sided phase noise spectral density}, \textcolor{mhs}{distinct from a single-sided spectrum, which is the sum of positive- and negative-frequency components; see Ref.~\cite{ieee2009ieee}}.}}\label{fig:performance}
\end{figure*}

\section{Experimental Results}

The sensitivity of a magnetometer can be determined from the response to a known applied magnetic field along with the measured noise. As AC magnetic fields are frequency-encoded in the oscillator magnetometer's $\approx\!5$~GHz output waveform, the measured phase noise sets the magnetic sensitivity of the device\textcolor{mhsnew}{; see} \textcolor{mhs}{SM Sec.~F~and~G~\cite{Note2}}\textcolor{mhsnew}{.} The oscillator magnetometer's single-sideband phase noise power spectral density $\mathcal{L}(f_m)$ is shown in Fig.~\ref{fig:performance}a. The device realizes a phase noise of -132.8 dBc/Hz at 10 kHz offset and -154.4 dBc/Hz at 100 kHz offset. 

Fitting Leeson's equation (Eqn.~\ref{eqn:leesonphasenoisevoltage}) to the oscillator's measured phase noise above 3 kHz gives $f_L = 600$ kHz, $F = 8$, and $f_c = 6.6$ kHz with the measured $P_s\approx3$ dBm. This value of $f_L=600$ kHz is in reasonable agreement with the value of $f_L = 755$ kHz expected from the FMR frequency discriminator's loaded linewidth.

To verify the device's response matches \textcolor{mhsnew}{that predicted} by Eqn.~\ref{eqn:response}, a sinusoidal magnetic field with rms amplitude $B_\text{sen}^\text{rms} = 2.12$ $\upmu$T is applied to the sensor, the angular frequency $\omega_m$ of this field is varied, and the carrier-normalized amplitude of the resulting sidebands is recorded. The measured data are in excellent agreement with the theoretical prediction of Eqn.~\ref{eqn:response}, as shown in Fig.~\ref{fig:performance}b. 

Having confirmed the device's frequency response is indeed governed by Eqn.~\ref{eqn:response}, the measured phase noise spectrum can be converted to a sensitivity spectrum by Eqn.~\ref{eqn:sensitivitynoleeson}, and the result is shown in Fig.~\ref{fig:performance}c.  As discussed previously, the sensitivity is expected to be approximately flat in the region between the amplifier noise corner at $f_c \approx 6.6$~kHz and the Leeson frequency $f_L \approx 600$~kHz. The measured data are consistent with this expectation; for AC signals of unknown phase we observe a minimum sensitivity of approximately 100~fT$/\sqrt{\text{Hz}}$ and a sensitivity below 200 fT$/\sqrt{\text{Hz}}$ over the band from 3 kHz to 1 MHz. 

\begin{figure*}[t]
\hspace{-3mm}
\begin{minipage}[b]{0.325\textwidth}
\begin{overpic}[height=1.82in]{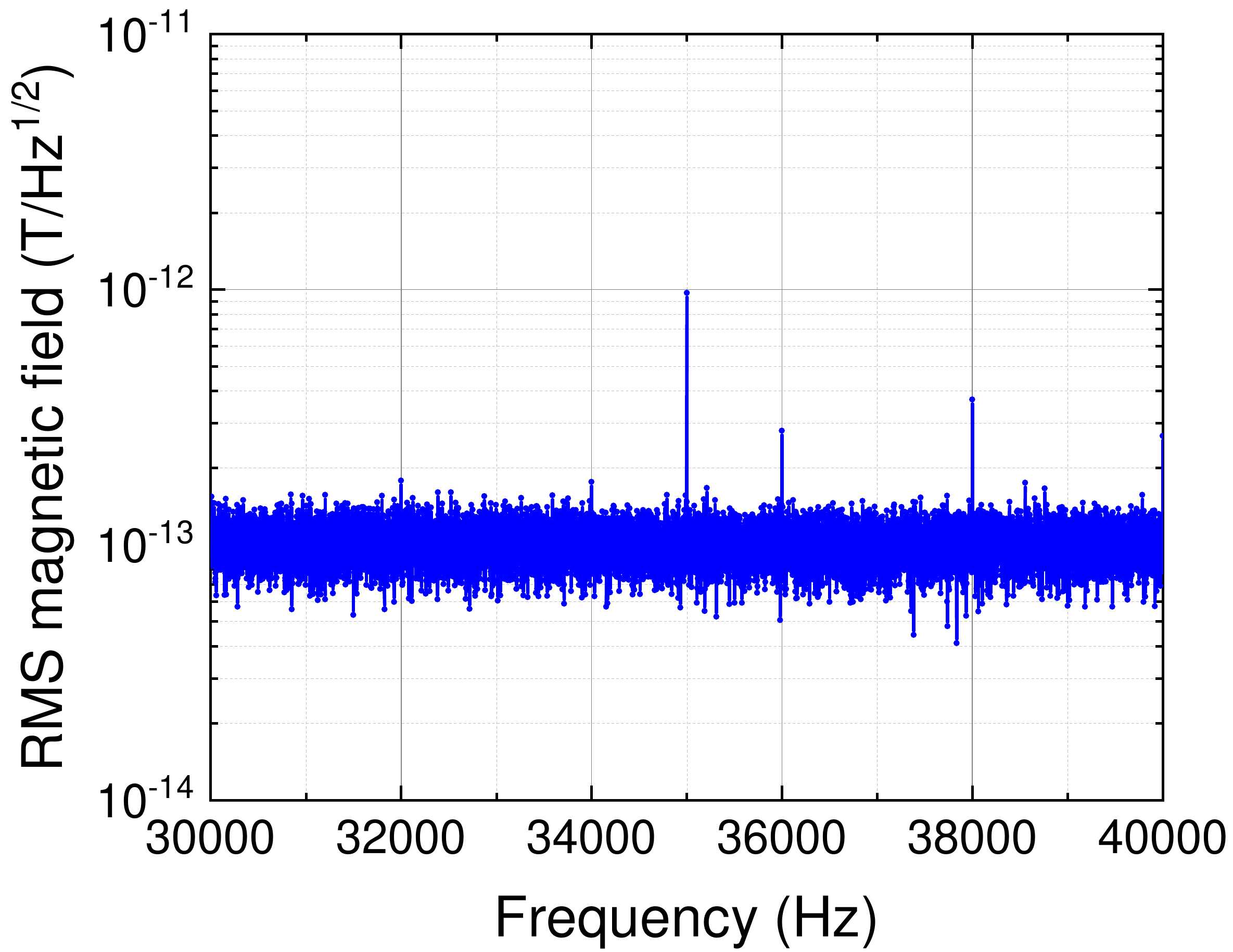} 
\put(0,78){\textbf{a)}}
\end{overpic}
\end{minipage}
\;\;
\begin{minipage}[b]{0.325\textwidth}
\begin{overpic}[height=1.82in]{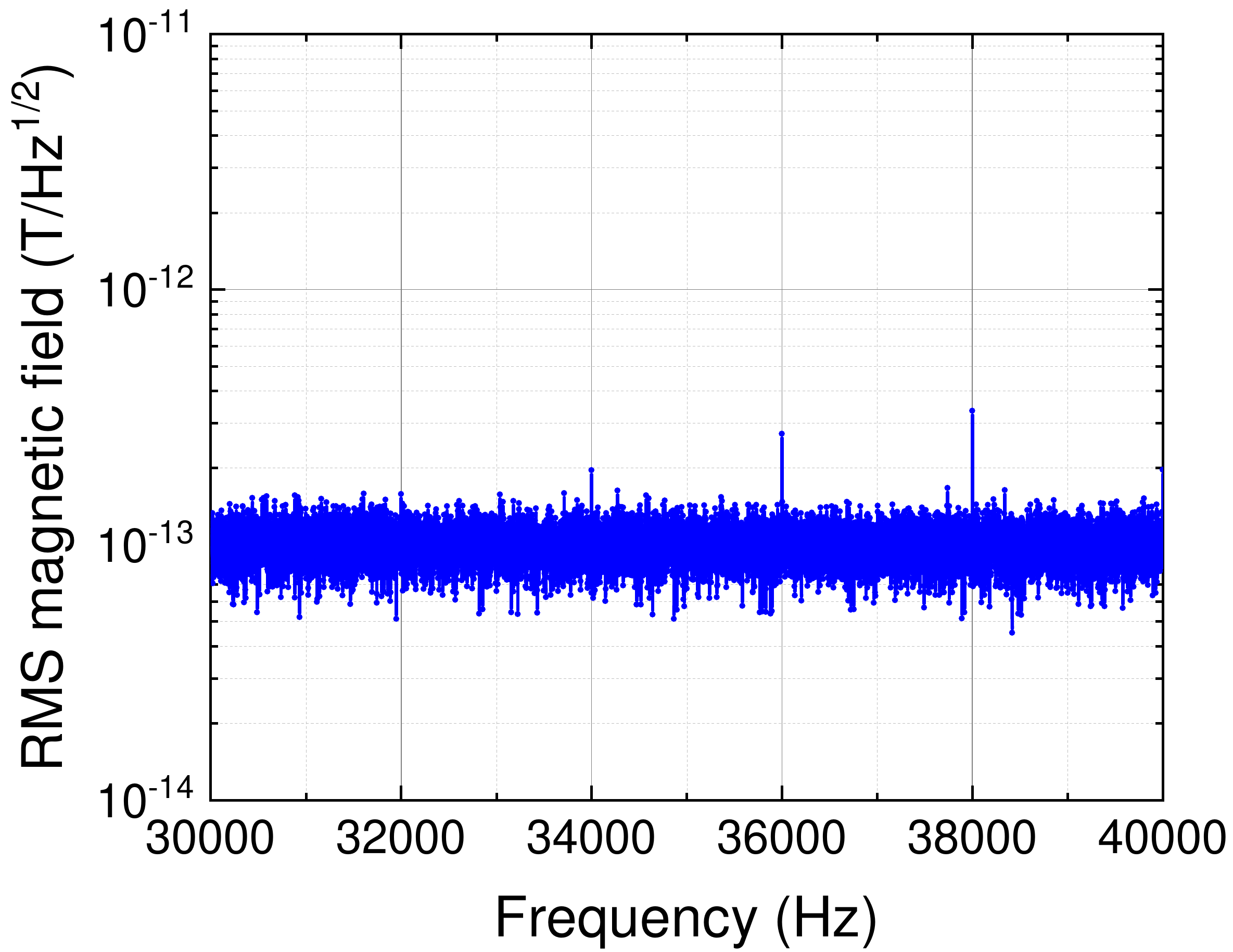} 
\put(-1,78){\textbf{b)}}
\end{overpic}
\end{minipage}
\begin{minipage}[b]{0.325\textwidth}
\begin{overpic}[height=1.8in]{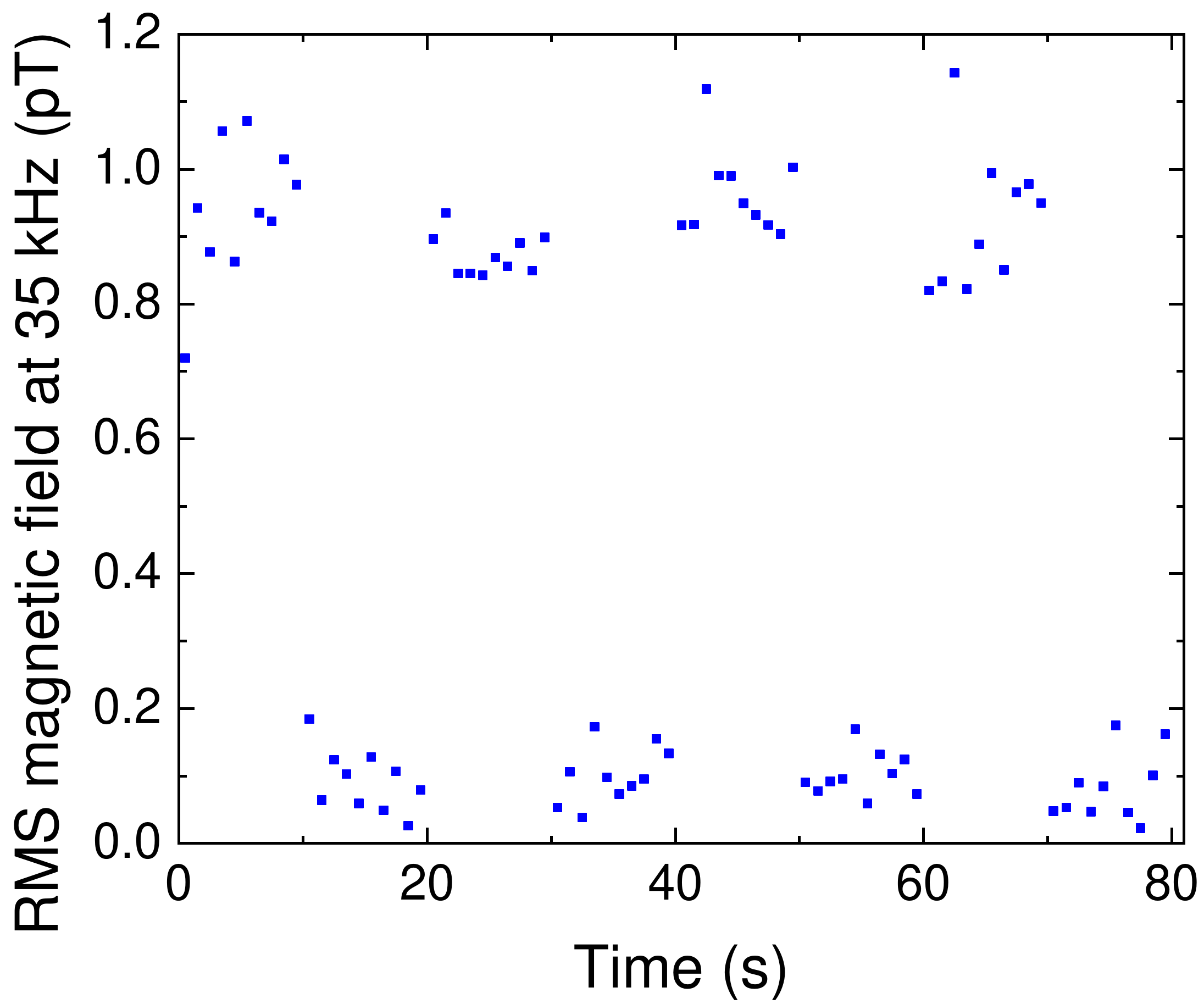} 
\put(-1,85){\textbf{c)}}
\end{overpic}
\end{minipage}
\caption{\textcolor{john}{\textbf{Magnetometer sensitivity determined from magnetic field time series.} (a) Single-sided magnetic field amplitude spectral density in 1 Hz bins with test field $B_\text{sen}^\text{rms}=0.9$ pT applied at 35 kHz. The spectrum is obtained by dividing a 10-s magnetic field time series into ten 1-s segments, computing the discrete Fourier transform for each segment, adding components at positive and negative frequencies in quadrature to convert to single-sided spectra, and rms-averaging the ten spectra together. (b) Single-sided magnetic field amplitude spectral density in 1 Hz bins without the test signal applied, obtained as in (a). (c) Value of the 35 kHz frequency bin calculated from 1-s of data per point as the $B_\text{sen}^\text{rms}=0.9$ pT signal is chopped on and off. As the device is tested in an unshielded environment, a \textcolor{allnew}{Tukey} window with $\alpha = 0.01$ prevents spectral leakage of low-frequency noise. This window is nearly rectangular, as $\alpha= 0$ and $\alpha=1$ correspond to rectangular and Hann windows respectively. The observed 100 fT/$\sqrt{\text{Hz}}$ sensitivity is consistent with that calculated from measured phase noise, shown in Fig.~\ref{fig:performance}c.}}\label{fig:mixerdemodulation}
\end{figure*}

\textcolor{john}{To operate the device as a practical magnetometer, the oscillator output is mixed down and digitized. The magnetic field time series is recovered from the digitized voltage waveform as described in SM Sec.~I~\cite{Note2}. To confirm device performance, a \textcolor{mhsnew}{35 kHz sinusoidal} test field $B_\text{sen}^\text{rms} = 0.9$ pT is \textcolor{mhsnew}{applied along} the sensor's z axis. The resultant amplitude spectral density with the test field on and off is shown in Figs.~\ref{fig:mixerdemodulation}a and \ref{fig:mixerdemodulation}b respectively\textcolor{mhs}{, and a time series of the 35 kHz signal size with the test field chopped on and off is shown in} Fig.~\ref{fig:mixerdemodulation}c. All data are consistent with the expected device response and a minimum sensitivity of 100~fT/$\sqrt{\text{Hz}}$. Supplemental Material Sec.~K \cite{Note2} details calibration of the test field.}

\section{Discussion}

\textcolor{john}{The device phase noise of -132.8 and -154.4 dBc/Hz at 10 and 100 kHz offset frequencies compares favorably to the lowest-phase-noise commercial YIG oscillators presently available~\cite{microlambda2022electromagnetic,teledyne2022tiny}. \textcolor{mhs}{The best commercial device we have measured achieves} -112 and -134 dBc/Hz at 10 and 100 kHz from its 5 GHz carrier frequency. \textcolor{mhs}{The improved performance} of our device is likely \textcolor{mhs}{mainly} attributable to a difference in $Q_L$; whereas we observe $f_L = 600 \;\text{kHz} \approx \frac{1}{2\pi}\frac{\omega_y}{2Q_L}$, the best commercial oscillator measured exhibits $f_L = 5.2$ MHz. This difference in $f_L$ should translate to an 18.8 dB improvement in phase noise at offset frequencies below $f_L$, similar to the observed difference of approximately 20 dB.}

The device demonstrated here provides the best AC sensitivity achieved to date for a solid-state quantum magnetometer~\cite{mitchell2020quantum,barry2016optical,wolf2015subpicotesla,eisenach2021cavity,fescenko2020diamond,schloss2018simultaneous,wilcox2022thermally}. \textcolor{mhsnew}{Among quantum magnetometers, this} sensitivity performance is surpassed only by cryogenic SQUID magnetometers and vacuum-based, optically-pumped vapor cell magnetometers. Additional sensitivity improvement may be attained by increasing the frequency response to magnetic fields or by decreasing the phase noise. For example, the frequency response could possibly be increased above $\gamma = 2\pi\times 28$ GHz/T in Eqn.~\ref{eqn:sphereresonance} \textcolor{mhsnew}{using} strong cavity coupling schemes~\cite{eisenach2021cavity}. Reference~\cite{krupka2020electrodynamic} describes such a scheme for a ferrimagnetic system with a predicted frequency response of $\approx 2\pi\times 500$~GHz/T. Cavity-enhanced ferrimagnetic oscillator magnetometers are currently under investigation and may be explored in future work.

On the other hand, methods to improve oscillator phase noise would also improve sensitivity and are well-established. Increasing the sustaining power $P_s$ is a common method to improve oscillator phase noise. However, this approach would likely improve phase noise only at frequencies above \textcolor{mhsnew}{the flicker} noise corner $f_c$, and $f_c$ itself may increase with larger values of $P_s$~\cite{boudot2012phase,rubiola2009phasebook}. Further, the maximum usable sustaining power is \textcolor{john}{presently believed to be limited} by instabilities arising from non-linear coupling of the uniform precession mode to spin-wave modes~\cite{suhl1956nonlinear,lax1962microwave}. \textcolor{mhs}{Under some conditions not far above current operating powers, we have seen indications that applying additional power to the YIG causes a binary change in the phase noise spectrum of the oscillator, with substantially deteriorated performance}.

Other approaches to improving overall phase noise might focus on the amplifier's additive phase noise. Amplifier-induced phase noise can be  mitigated using oscillator-narrowing techniques such as Pound-Drever-Hall locking~\cite{pound1946electronic,drever1983laser,galani1984analysis,locke2008design,fluhr2016characterization},  carrier suppression interferometric methods~\cite{ivanov1996ultra,ivanov1998applications,ivanov2006low,ivanov2009low,rubiola2002advanced,gupta2004high}, careful design~\cite{walls1997origin,pikal1997guidelines} and other approaches~\cite{rubiola2009phasebook}. However, even in the ideal case, oscillator-narrowing techniques cannot reduce the oscillator's phase noise to the thermal noise limit of -177 dBm/Hz expected in the absence of Leeson gain. While lowering the Leeson frequency $f_L$ will improve phase noise performance, the noise gain introduced by the Leeson effect appears to be fundamental to the oscillator architecture, as  discussed in \textcolor{mhs}{SM Sec.~G \cite{Note2}}.

In conclusion, the magnetometer design reported here offers a unique combination of state-of-the-art sensitivity with realistic prospects for improvement, high dynamic range, compactness, and low power requirements. These advantages could drive widespread adoption of similar quantum sensing devices in the near future. The oscillator architecture can be adapted to simplify high-performance ensemble sensing with a range of quantum materials and in a variety of sensing applications, such as sensing of electric fields~\cite{michl2019robust,chen2017high,block2020optically}, temperature~\cite{kucsko2013nanometre}, or pressure~\cite{hsieh2019imaging}.

\section{Acknowledgements}
The authors acknowledge Peter F. Moulton, Kerry A. Johnson, Liam J. Fitzgerald, and Erik R. Eisenach for helpful discussions. \textcolor{mhs}{This research was developed with funding from the Defense Advanced Research Projects Agency} and the Under Secretary of Defense for Research and Engineering under Air Force Contract No. FA8702-15-D-0001. The views, opinions, and/or findings expressed are those of the authors and should not be interpreted as representing the official views or policies of the Department of Defense or the U.S. Government. \textcolor{newall}{R.A.I. and M.H.S. contributed equally to this work.}


\bibliographystyle{apsrev4-2_custom}
\bibliography{ferrimagneticbiblio_v6}



\cleardoublepage

\textcolor{mhs}{\section*{Supplemental Material For ``A ferrimagnetic oscillator magnetometer''}}

\subsection{Variable names and symbols}
A list of variables and constants referenced in this work is given in Table~\ref{tab:varnames}. Where possible we adopt the notation of Ref.~\cite{rubiola2009phasebook}.

\begin{table*}[h t b p]  
\centering
\caption{Partial list of variables} 
\centering 
\begin{tabular}{l l c c } 
\hline\hline   
Name & Symbol & Approx. value & Units  \\
\hline   
Gyromagnetic ratio & $g_e$ & $ 2$ & unitless \\
Bohr magneton & $\mu_B$ & $9.274\times10^{-24}$ & J/T \\
Vacuum permeability & $\mu_0$ & $1.257\times 10^{-6}$ & H/m \\
Boltzmann constant & $k_B$ & $1.381\times 10^{-23}$ & J/K \\
Gyromagnetic ratio & $\gamma$ & $2\pi \times 28\times 10^9$ & rad/(s$\cdot$T) \\
System temperature & $T$ & $ 300$ & K \\
Oscillator's measured wideband noise factor & $F$ & 8 & unitless \\
$1/f$ flicker noise corner & $f_c$ & $6.6\times10^3$ & Hz\\
Amplifier input (sustaining) power & $P_s$ & 0.002 & W \\
Intrinsic linewidth & $\kappa_0$ & $2\pi \times 7.90\times 10^5$ & rad/s\\
Input coupling rate & $\kappa_1$ & $2\pi \times 3.15\times 10^5$ & rad/s \\
Output coupling rate & $\kappa_2$ & $2\pi \times 4.05\times 10^5$& rad/s \\
YIG resonant frequency & $\omega_y$ & $2\pi \times 5\times 10^9$ & rad/s\\
MW drive frequency & $\omega_d$ & $2\pi \times 5\times 10^9$ & rad/s\\
Oscillator carrier frequency & $\omega_c$ & $2\pi \times 5\times 10^9$ & rad/s\\
Intermediate frequency & $\omega_i$ &  - & rad/s\\
Loaded linewidth & $\kappa_L = \kappa_0\!+\!\kappa_1\!+\!\kappa_2$ & $2\pi \times 1.51\times 10^6$& rad/s \\
Loaded quality factor & $Q_L = \omega_y/\kappa_L$ & 3300 & unitless \\
Unloaded quality factor & $Q_0 = \omega_y/\kappa_0$ & 6300 & unitless \\
Leeson frequency & $f_L = \kappa_L/(2\times 2\pi)$ & $\sim 6.0\times 10^5$ & Hz \\
S-parameters & $S_{11},S_{12},S_{21},S_{22}$ & - & unitless \\
Total magnetic field & $\mathbf{B}=\mathbf{B_0}+\mathbf{B}_\text{sen}$ & - & T\\
Bias magnetic field & $\mathbf{B_0} = B_0 \hat{z}$ & 0.1780 & T \\
Test sensing magnetic field & $\mathbf{B}_\text{sen}(t)$ & - & T \\
Test sensing magnetic field rms amplitude & $B_\text{sen}^\text{rms}$ & $2.12\times10^{-6}$ & T \\
Test sensing magnetic field angular frequency & $\omega_m = 2\pi f_m$ & - & rad/s \\
Test sensing magnetic field frequency & $f_m=\omega_m/(2\pi)$ & - & Hz \\
Demagnization factors & $N_x,N_y,N_z$ & $\frac{1}{3},\frac{1}{3},\frac{1}{3}$ & unitless \\
Saturation magnetization & $M_s$ & $1.42\times10^5$ & A/m \\
Single-sideband phase noise power spectral density & $\mathcal{L}(f_m)$ & - & dBc/Hz\\
Single-sideband phase noise amplitude spectral density & $\mathcal{L}^\frac{1}{2}(f_m)$ & - & dBc/$\sqrt{\text{Hz}}$\\
Time & $t$ & - & s \\
Oscillator voltage waveform & $v(t)$ & - & V\\
Oscillator voltage waveform amplitude & $V_0$ & - & V\\
Oscillator instantaneous phase & $\phi(t)$ & - & rad \\
Oscillator additive phase (noise or signal) & $\varphi(t)$ & - & rad\\
Oscillator additive amplitude noise & $\alpha(t)$ & - & unitless \\
Line impedance & $Z_0$ & - & $\Omega$ \\
Transformer turn ratios & $n_1,n_2$ & - & unitless \\
Magnetic sensitivity at frequency $f_m$ & $\eta(f_m)$ & - & T/$\sqrt{\text{Hz}}$\\
YIG sphere volume & $V_y$ & $5.2\times 10^{-10}$ & m$^3$ \\
\hline 
\end{tabular}\label{tab:varnames}
\end{table*}  

\subsection{Qualitative details of oscillator phase noise}\label{sec:phasenoisedetails}

The phase noise predicted by Leeson's equation can be interpreted as follows: Additive phase noise in the oscillator loop below the Leeson frequency $f_L$ passes through the FMR frequency discriminator and is regeneratively amplified into the oscillator's output waveform. Consequently, the system's aggregate phase noise depends strongly on both the Leeson frequency $f_L$ and additive phase noise from components within the oscillator loop. The sustaining amplifier, with both flicker phase noise below \textcolor{mhsnew}{approximately} $f_c$ and wideband phase noise characterized by the amplifier noise factor $F$~\cite{boudot2012phase}, is a primary contributor to this in-loop phase noise. In contrast, additive phase noise introduced by components outside the oscillator loop, such as buffer amplifiers, mixers, and digitizers, \textcolor{mhs}{experiences} no such regenerative gain~\cite{rubiola2009phasebook} and \textcolor{mhs}{contributes} much less to the total \textcolor{mhsnew}{oscillator} phase noise.

\subsection{Additional details of YIG}\label{sec:additionaldetailsYIG}

For a single magnetic domain at absolute zero, YIG exhibits a net magnetic moment equal to that of one Fe$^{3+}$ atom per every 20 lattice atoms~\cite{stancil2009spin}, resulting in a polarized electron spin density of $2.1\times 10^{22}/$cm$^3$. Magnetization at room temperature retains 72$\%$ of the zero-temperature magnetization~\cite{anderson1964molecular}, equal to a polarized electronic spin density of $1.5\times10^{22}$/cm$^3$. For comparison, typical paramagnetic spin systems exhibit spin densities within a few orders of magnitude of $10^{17}$/cm$^3$~\cite{barry2020sensitivity}, while alkali vapor cells operate with a spin density \mbox{$\sim 10^{13}$/cm$^3$}~\cite{kominis2003subfemtotesla}.

Magnetometer performance depends upon localizing the ferrimagnetic resonance with precision, accuracy and speed. The precision with which the FMR resonance can be localized, and therefore the ambient magnetic field determined, depends on the FMR intrinsic linewidth $\kappa_0$. Single-crystal YIG exhibits the lowest linewidth of all known ferromagnetic or ferrimagnetic materials, with highly polished YIG spheres~\cite{lecraw1958ferromagnetic} exhibiting measured linewidths of $2\pi \times 560$ kHz or below~\cite{roschmann1981annealing,pacewicz2019rigorous,lecraw1959surface}. The material employed in this work exhibits a FWHM linewidth of $\kappa_0 \approx$ $2\pi \times 790$ \textcolor{mhsnew}{kHz at} $\omega_y \textcolor{mhsnew}{\approx}$ $2\pi \times 5$ GHz.


Minimal values of $\kappa_0$ occur when the YIG crystal's magnetic domains are uniformly oriented, which is achieved by applying an external bias magnetic field with sufficient strength to saturate the magnetization~\cite{lax1962microwave}. \textcolor{john}{For pure YIG an external bias field $B_0 \approx 0.178$~T is more than sufficient to saturate the magnetization.} Device operation in the saturated magnetization regime is important to ensure the ferrimagnetic resonance displays a constant response to changes in the externally applied magnetic field, namely $\frac{d\omega_y}{dB} = \gamma$~\cite{lax1962microwave,helszajn1985yig}.

The high spin density and strong coupling between spins, which prevents deleterious broadening, allows \mbox{$\sim$ 100 fT/$\sqrt{\text{Hz}}$} sensitivities to be achieved using crystal volumes $\lesssim \! 1$ mm$^3$. Magnetic gradients can compromise sensitivity when the gradient broadening becomes comparable to the intrinsic linewidth. Assuming an intrinsic linewidth of $\kappa_0 = 2\pi \times 1$~MHz results in a gradient tolerance $\sim$ 0.4 mT/cm before substantial degradation of sensor performance is expected. This gradient tolerance compares favorably to the $\sim 30$ nT/cm gradient tolerance characteristic of alkali vapor magnetometers, which typically have sample volume length scales $\sim 10 \times$ larger and intrinsic linewidths $\sim 10^3\times$ smaller than the device reported here.

\subsection{Equivalent circuit and S-parameters for a YIG transmission filter}\label{sec:equivcircuit}
 
The equivalent circuit for a \textcolor{mhs}{transmission-topology} YIG with orthogonal coupling loops (e.g., two \textcolor{john}{coupling} loops in planes rotated from each other by $\pi/2$, the geometry of this device) is shown in Fig. \ref{fig:YIGequivalentcircuit} and closely resembles the equivalent circuit for a series RLC circuit with inductive coupling~\cite{carter1965side,Tanbakuchi1987broadband,helszajn1985yig,montgomery1987principles}. The only difference is that the gyrator action of the YIG introduces a direction-dependent (i.e. non-reciprocal) phase shift: $S_{21}$ is retarded by $\pi/2$ while $S_{12}$ is advanced by $\pi/2$. The $S$ parameters describing this system are 
\begin{align}\label{eqns:sparameter1}
S_{11} &= 1-\frac{\kappa_{1}}{i(\omega_d-\omega_y) +\frac{\kappa_{0}+\kappa_{1}+\kappa_{2}}{2}},\\\label{eqns:sparameter2}
S_{21} &= \frac{\sqrt{\kappa_{1}\kappa_{2}}}{i(\omega_d-\omega_y) +\frac{\kappa_{0}+\kappa_{1}+\kappa_{2}}{2}}e^{-i\frac{\pi}{2}},\\ \label{eqns:sparameter3}
S_{12} &= \frac{\sqrt{\kappa_{1}\kappa_{2}}}{i(\omega_d-\omega_y) +\frac{\kappa_{0}+\kappa_{1}+\kappa_{2}}{2}}e^{i\frac{\pi}{2}},  \\ \label{eqns:sparameter4}
S_{22} &= 1-\frac{\kappa_{2}}{i(\omega_d-\omega_y) +\frac{\kappa_{0}+\kappa_{1}+\kappa_{2}}{2}},\
\end{align}
where $\omega_d$ is the drive frequency, $\omega_y$ is the FMR frequency, $\kappa_{0}=\frac{R}{L}$ is the intrinsic YIG linewidth, $\kappa_{1}= \frac{Z_0}{n_1^2 L}$ is the input coupling rate, and $\kappa_{2}= \frac{Z_0}{n_2^2 L}$ is the output coupling rate; the resistance $R$, the inductance $L$, the line impedance $Z_0$, and the transformer turns ratios $n_1$ and $n_2$ are parameters of the RLC model. The parameters $\omega_d$, $\omega_y$, $\kappa_0$, $\kappa_1$, and $\kappa_2$ are all in angular units. The intrinsic linewidth of the FMR filter is given by $\kappa_0 = \frac{\omega_y}{Q_0}$ where $Q_0$ is the unloaded quality factor of the YIG sphere, extracted from measurements performed by sweeping $\omega_d$ while $B_0$ is fixed. The intrinsic linewidth can also be determined by sweeping the value of $B_0$ for a fixed value of $\omega_d$, and literature values of YIG linewidths are often given in magnetic field units rather than frequency units~\cite{lax1962microwave}. Eqns. \ref{eqns:sparameter1}-\ref{eqns:sparameter4} are valid only near resonance, both because of the inclusion of the ideal transformer and because the equations have been symmetrized about the resonance frequency.

While changing the coupling loop diameters allows adjustment of $\kappa_1$ and $\kappa_2$, the value of $\kappa_0$ is less easily varied and depends in part on the chemical purity of the YIG material, the sphere's surface finish, and the uniformity of the bias magnetic field.

In practice, we found that the requirement of orthogonal coupling loops is not stringent; minor twisting and positional variation of the coupling loops did not produce problematic off-resonant coupling.

\subsection{Determination of intrinsic linewidth}\label{determiningintrinsiclinewidth}

\begin{figure}[t]
\centering
\includegraphics[width=3.4in]{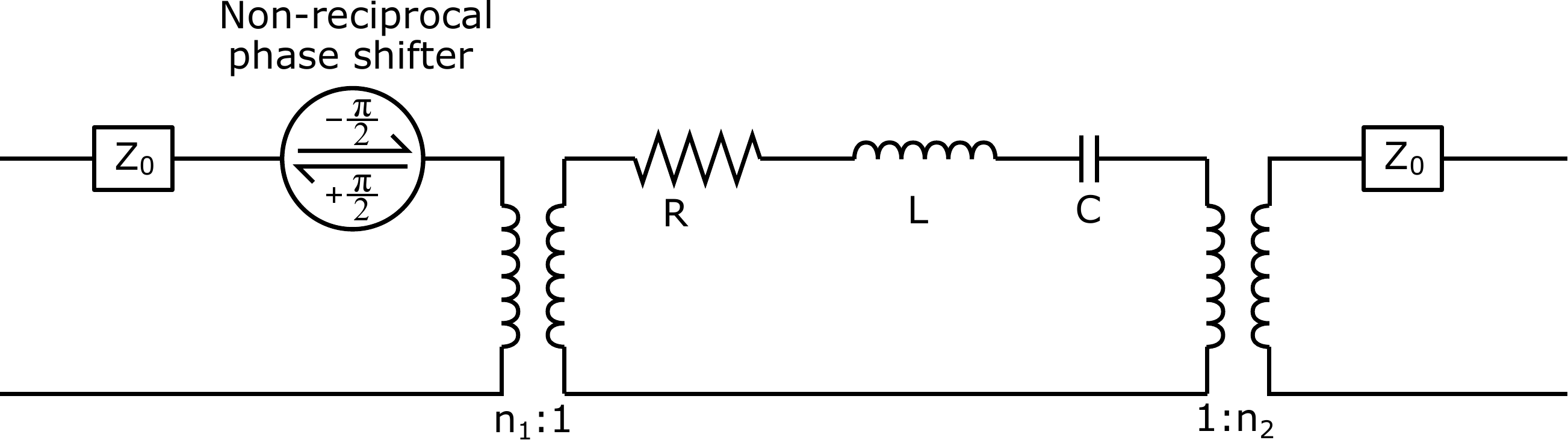}
\caption{The YIG transmsision filter can be modelled as a series RLC circuit with idealized inductive coupling and a non-reciprocal $\pi/2$ phase shifter to account for the gyrator action of the ferrimagnetic resonance.}\label{fig:YIGequivalentcircuit}
\end{figure}

We wish to determine the intrinsic linewidth $\kappa_0$ of the uniform mode ferrimagnetic resonance from S-parameter data measured by a vector network analyzer (VNA) on a two-port YIG transmission filter, as shown in Fig~\ref{fig:YIGequivalentcircuit}. From Eqns. \ref{eqns:sparameter2} and \ref{eqns:sparameter3}, we determine the maximum and minimum of the power transmission and reflection, respectively,
\begin{align}
|S_{11}|^2_\text{min} &= \left(1-\frac{\kappa_{1}}{ ( \frac{\kappa_{0}+\kappa_{1}+\kappa_{2}}{2})}\right)^2, \\
|S_{21}|^2_\text{max} &=  \frac{\kappa_{1}\kappa_{2}}{ \left(\frac{\kappa_{0}+\kappa_{1}+\kappa_{2}}{2}\right)^2}.
\end{align}
The value of the loaded linewidth, 
\begin{equation}
\kappa_L = \kappa_{0}+\kappa_{1}+\kappa_{2},
\end{equation}
can be determined from the distance \textcolor{mhsnew}{in} frequency between the points where $|S_{21}|^2$ is reduced by 3~dB from its peak value. The system can then be solved for $\kappa_{0}$, $\kappa_{1}$, and $\kappa_{2}$, as there are three equations and three unknowns. Thus, we have
\begin{align}
\kappa_{0} &= \frac{\kappa_L}{2} \left(1+\sqrt{|S_{11}|^2_\text{min}}-\frac{|S_{21}|^2_\text{max}}{1-\sqrt{|S_{11}|^2_\text{min}}}\right), \\
\kappa_{1} &= \frac{\kappa_L}{2} \left( 1-\sqrt{|S_{11}|^2_\text{min}} \right),\\
\kappa_{2} &= \frac{\kappa_L}{2}  \left( \frac{|S_{21}|^2_\text{max}}{1-\sqrt{|S_{11}|^2_\text{min}}} \right).
\end{align}

\subsection{Ferrimagnetic oscillator magnetometer theory of operation}\label{sec:operation}

For the uniform mode of ferrimagnetic resonance in a spherical sample with saturated magnetization, the time derivative of the instantaneous phase $\phi(t)$ obeys
\begin{equation}\label{eqn:dphase}
\frac{d\phi(t)}{dt} = \gamma B(t),
\end{equation}
where $B(t)$ is the externally applied magnetic field and $\gamma = \frac{g_e \mu_B}{\hbar}$ (with the electron g-factor $g_e \approx 2$, the Bohr magneton $\mu_B$, and the reduced Planck's constant $\hbar$, so that $\gamma \approx 2\pi\!\times\! 28$ GHz/T\textcolor{mhsnew}{)}. For the purpose of this discussion, we neglect crystal anisotropy (see \textcolor{mhsnew}{SM Sec.~L}) which introduces higher-order terms into Eqn.~\ref{eqn:dphase}. 


The precessing magnetization of the sphere described by Eqn.~\ref{eqn:dphase} inductively couples to the output coupling loop, producing a voltage signal which is then amplified by the sustaining amplifier and inductively coupled back to the precessing magnetization; this closed \textcolor{john}{feedback} loop produces sustained self-oscillation \textcolor{mhsnew}{as described in the main text}. The oscillator output voltage is then $v(t) = V_0 \cos[\phi(t)]$ where $V_0$ represents the oscillator's voltage amplitude \footnote{Since the oscillating voltage at any point in the oscillator loop has a fixed phase relative to the magnetization, we do not distinguish between the phase of the magnetization and the phase of the oscillator's voltage output}. 

The oscillator phase $\phi(t)$ is continuous in time and given by
\begin{equation}
\phi(t) = \int_0^t \frac{d\phi(\tau)}{d\tau} d\tau = \int_0^t \gamma B(\tau) d\tau,
\end{equation}
where we have set $\phi(t=0) \equiv 0$. The total magnetic field $\mathbf{B}(t)$ seen by the sensor is the vector sum of a static field $\mathbf{B}_0=B_0 \mathbf{\hat{z}}$ \textcolor{mhsnew}{created} by the permanent \textcolor{mhsnew}{magnets} and the ambient field $\mathbf{B}_\text{sen}(t)$ external to the sensor. For simplicity we assume $\mathbf{B}_\text{sen}(t)$ lies along $\mathbf{\hat{z}}$, but the case of arbitrary $\mathbf{B}_\text{sen}(t)$ is easily worked out (see \textcolor{mhsnew}{SM~Sec.~N}). The value of $B_0$ is assumed to exhibit only slow temporal variation (\textcolor{mhsnew}{e.g.,} due to thermal drift of the magnets or vibration of the mechanical structure holding the magnets) on time scales below the frequencies of interest, so that $B_0$ can be treated as constant. Then the total magnetic field seen by the ferrimagnetic sphere is $B(t) = B_0 + B_\text{sen}(t)$, allowing the oscillator phase to be expressed as
\begin{equation}\label{eqn:oscillatorphasevsBfield}
\phi(t) = \int_0^t \gamma\big[B_0 + B_\text{sen}(\tau)\big] d\tau.
\end{equation}
An arbitrary real waveform $B_\text{sen}(t)$ can be decomposed into its Fourier series as
\begin{equation}
B_\text{sen}(t) = \frac{a_0}{2} + \sum_{n=1}^\infty a_n \cos(\omega_n t) + \sum_{n=1}^\infty b_n \sin(\omega_n t).
\end{equation} 
For simplicity we assume $B_\text{sen}(t)$ consists of a single spectral component such that $B_\text{sen}(t) = \sqrt{2}B_\text{sen}^\text{rms}\big[\cos \omega_m t \big]$,
where $\omega_m$ is the angular frequency of the \textcolor{mhsnew}{ambient} external magnetic field and $B_\text{sen}^\text{rms}$ is the rms field amplitude. With this simplification, the oscillator phase is
\begin{align*}
\phi(t) &= \int_0^t \gamma \big[B_0 + \sqrt{2}B_\text{sen}^\text{rms}\big[\cos \omega_m \tau \big]\big] d\tau \\
 &= \gamma B_0 t + \sqrt{2}\frac{\gamma B_\text{sen}^\text{rms}}{\omega_m}\big[\sin \omega_m t \big].
\end{align*}
The oscillator waveform is then 
\begin{equation}\label{eqn:hardformwaveform}
v(t) = V_0 \cos \big[\gamma B_0 t +\sqrt{2}\frac{\gamma B_\text{sen}^\text{rms}}{\omega_m}   \sin[ \omega_m t]\big].
\end{equation}
\textcolor{john}{We now outline the steps to manipulate Eqn.~\ref{eqn:hardformwaveform} to a more convenient form. A similar derivation can be found in Ref.~\cite{robins1984phase}. From the Jacobi-Anger expansion \cite{cuyt2008handbook}, we can derive the Bessel function identity
\begin{equation}\label{eqn:besselidentity}
\cos(\omega_c t + \beta \sin(\omega_m t)) = \sum_{k=-\infty}^\infty J_k(\beta) \cos\big[(\omega_c+k\omega_m)t\big].
\end{equation}
For $k \geq 0$, $J_k(\beta)$ may be expressed to leading order in $\beta$ as \cite{cuyt2008handbook}
\begin{equation}
J_k(\beta) \approx \left(\frac{\beta}{2}\right)^k \frac{1}{\Gamma(k + 1)},
\end{equation}
where $\Gamma$ denotes the gamma function, and for integer $k$ the Bessel functions satisfy \cite{cuyt2008handbook}
\begin{equation}
J_{-k}(x) = (-1)^k J_k(x).
\end{equation}
For $\beta \ll 1$, only terms with small $|k|$ contribute. The series in Eqn.~\ref{eqn:besselidentity} can then be approximated using only the $k = -1, 0, 1$ terms, giving
\begin{align}\label{eqn:threeterms}
\cos(\omega_c t + \beta \sin(\omega_m t)) & \approx \cos(\omega_c t) \nonumber \\+ \frac{\beta}{2}\cos[(\omega_c + &\omega_m)t]
- \frac{\beta}{2}\cos[(\omega_c - \omega_m)t)]
\end{align}
Comparing Eqns.~\ref{eqn:hardformwaveform} and \ref{eqn:threeterms}, we identify $\omega_c = \gamma B_0$ and $\beta = \sqrt{2}\frac{\gamma B_{\mathrm{sen}}^{\mathrm{rms}}}{\omega_m}$. Then if $\gamma B_\text{sen}^\text{rms} \ll \omega_m$, we have $\beta \ll 1$ and Eqn.~\ref{eqn:hardformwaveform} can be rewritten as}
\begin{align}
v(t) & \textcolor{mhs}{\approx} V_0 \bigg[ \cos[\gamma B_0 t] + \frac{\gamma B_\text{sen}^\text{rms}}{\sqrt{2}\omega_m}\cos[(\omega_c+\omega_m)t]\\ \nonumber
& \hspace{1cm} - \frac{\gamma B_\text{sen}^\text{rms}}{\sqrt{2}\omega_m}\cos[(\omega_c-\omega_m)t]\bigg].
\end{align}
For external fields satisfying $\gamma B_\text{sen}^\text{rms} \ll \omega_m$, the B-field FM-modulation results in two antisymmetric sidebands at $\pm \omega_m$, each with amplitude $\gamma B_\text{sen}^\text{rms}/(\sqrt{2}\omega_m)$. For example, a 1 pT RMS magnetic field at 100 kHz produces two sidebands each with power -134 dBc.

\subsection{YIG magnetometer noise}\label{sec:yigmagnetometernoise}

As detailed in the preceeding section, a single-frequency AC magnetic field applied to the sensor results in frequency modulation of the oscillator's carrier at angular frequency $\omega_m$. In the frequency domain, this modulation manifests as two sidebands offset by $\pm \omega_m$ from the oscillator's carrier frequency, each with a carrier-normalized amplitude of
\begin{equation}
s=\frac{\gamma B_\text{sen}^\text{rms}}{\sqrt{2}\omega_m}.
\end{equation} 
Magnetic field detection then reduces to resolving these two sidebands from the oscillator's measured phase noise. The magnetic sensitivity $\eta (f_m)$ may be written as a ratio between the phase noise \textcolor{mhsnew}{amplitude} spectral density $\mathcal{L}^\frac{1}{2}(f_m)$ and the signal due to these FM sidebands, each with carrier-normalized amplitude $s$. Thus the sensitivity is
\begin{equation}\label{eqn:sensitivitysignalandphasenoiseappendix}
\eta(f_m) = \frac{\sqrt{2}f_m}{\gamma / (2\pi)} \times  \mathcal{L}^\frac{1}{2}(f_m),
\end{equation}
where $\mathcal{L}^\frac{1}{2}(f_m)$ is the single-sided phase noise spectral density of the oscillator. With optimal synchronous detection of a magnetic field with known phase, the sensitivity is improved by $\sqrt{2}$ over that expected from Eqn.~\ref{eqn:sensitivitysignalandphasenoiseappendix}, which assumes the phase of the B field is unknown. We also note that for realistic oscillators the phase noise power spectral density is symmetric about the carrier, and thus there is no improvement to be gained by processing both the upper and lower sideband (see \textcolor{mhsnew}{SM~Sec.~J}).

To predict the frequency dependence of the sensitivity, we can apply Leeson's model of oscillator phase noise to Eqn.~\ref{eqn:sensitivitysignalandphasenoiseappendix}. Leeson's equation 
for the single-\textcolor{mhsnew}{sideband} phase noise of an oscillator as a function of the offset frequency $f_m$ from the carrier \textcolor{mhsnew}{is} \cite{rhea2010discrete} 
\begin{equation}\label{eqn:leesonphasenoisevoltagesupplement}
\mathcal{L}^\frac{1}{2}(f_m) = \sqrt{\frac{1}{2}\left[\frac{f_L^2}{f_m^2}+1\right] \left[ \frac{f_c}{f_m} +1 \right] \left[\frac{F k_B T}{P_s}\right]},
\end{equation}
where $f_L \equiv \frac{1}{2}\frac{\kappa_L}{2\pi}$ denotes the Leeson frequency \cite{rubiola2009phasebook}, $f_c$ is the $1/f$ flicker noise corner~\cite{rubiola2009phasebook,boudot2012phase,robins1984phase}, $P_s$ is the input power to the sustaining amplifier, $T$ is the temperature, $k_B$ is Boltzmann's constant, and $F$ \textcolor{john}{is the oscillator's measured wideband} noise factor. We note that $\kappa_L$ as used in this work is an angular frequency FWHM while $f_L$ is a non-angular frequency half-width. Combining Eqns.~\ref{eqn:sensitivitysignalandphasenoiseappendix} and \ref{eqn:leesonphasenoisevoltagesupplement} \textcolor{mhsnew}{yields} an expected sensitivity of
\begin{equation}\label{eqn:fullsensitivityappendix}
\eta(f_m) = \frac{\sqrt{2}f_m}{\gamma/(2\pi)} \sqrt{\frac{1}{2}\left[\frac{f_L^2}{f_m^2}+1\right] \left[ \frac{f_c}{f_m} +1 \right] \left[\frac{F k_B T}{P_s}\right]}. 
\end{equation}
\textcolor{mhsnew}{Equation} \ref{eqn:fullsensitivityappendix} \textcolor{mhsnew}{suggests that best sensitivity will occur for} frequencies $f_m$ satisfying $f_c < f_m < f_L$. In this region, both the signal and the noise scale as $\approx\! 1/f_m$, resulting in an approximately flat frequency response. For the results reported here, we find $f_c = 6.6$ kHz and $f_L \approx 600$ kHz, and the device's best sensitivity is observed between those two frequencies as expected, as shown in Fig.~\ref{fig:performance}c. At frequencies below $f_c$ or above $f_L$, sensitivity is \textcolor{mhsnew}{reduced.} At frequencies near or below $f_c$, the flicker noise of the amplifier (as well as other effects such as thermal drift of the ferrimagnetic resonance or vibration) increases the oscillator phase noise relative to the signal.  For frequencies near or above the Leeson frequency $f_L$, sensitivity is compromised because the phase noise amplitude spectral density is independent of $f_m$ while the signal response \textcolor{mhsnew}{continues to decrease} as $f_m$ increases.

If $f_m$ additionally satisfies $f_c \ll f_m \ll f_L$, Eqn. \ref{eqn:fullsensitivityappendix} \textcolor{mhsnew}{simplifies} to
\begin{equation}\label{eqn:senstivitysupersimple}
\eta \approx \frac{1}{2} \frac{\kappa_L}{\gamma}\sqrt{\frac{F k_B T}{P_s}}.
\end{equation}
We note that setting $F=1$ in Eqn.~\ref{eqn:senstivitysupersimple} yields a sensitivity equivalent to that of an idealized (that is, thermal noise limited and with ideal amplifier), optimally-coupled ($\kappa_{1}=\kappa_2=\kappa_0/2$, assuming an ideal amplifier \cite{gronefeld2018ultralow}) transmission interferometer.

The above discussion reflects what appear to be fundamental limits of oscillator phase noise. Quantitatively the Leeson effect~\cite{leeson1966simple,rubiola2009phasebook} dictates that
\begin{equation}\label{eqn:leesoneffect}
S_\varphi (f_m) = \left[ 1+\frac{f_L^2}{f_m^2}\right] S_\psi(f),
\end{equation}
where $S_\psi(f_m)$ is the \textcolor{mhsnew}{single}-sided power spectral density of additive phase shifts inside the oscillator loop, and $S_\varphi(f_m)$ denotes the \textcolor{mhsnew}{single}-sided power spectral density of the oscillator's output phase noise. As thermal noise sets a lower bound on $S_\psi(f_m)$ and this noise is effectively enhanced in $S_\varphi(f_m)$ for frequencies below $f_L$, an oscillator would appear to be forbidden from reaching the naive thermal phase noise limit \textcolor{mhsnew}{[$\mathcal{L}(f_m) = -177$~dBm/Hz, equivalent to $S_\varphi (f_m) = -174$~dBm/Hz]} for frequencies below $f_L$, regardless of any oscillator narrowing techniques that may be used. The same limits are observed in inferometric frequency discriminators~\cite{ivanov1998applications}.

\subsection{\textcolor{mhs}{Tip angle}}\label{sec:tiltangle}

\textcolor{john}{The RF magnetic field $\mathbf{B}_\text{rf}$ tips the precessing magnetization away from the applied DC magnetic field $\mathbf{B}_0$. Given the 11 dBm of MW power applied to the YIG sphere, $B_\text{rf} \approx 2 \times 10^{-6}$ T is estimated from the known geometry. The tip angle is then calculated using
\begin{equation}
\theta_\text{tip} = \arccos\left[\frac{1}{1+\frac{1}{4}\gamma^2 B_\text{rf}^2 T_1 T_2}\right].
\end{equation}
For $T_2 = \frac{1}{\pi\times 790 \text{ kHz}} = 400$ ns, and approximating $T_1 = T_2/2$, we expect $\theta_\text{tip} \approx 0.1$ radians.}

\subsection{Demodulation}\label{sec:demodulation}

The magnetic field $B(t)$ applied to the sensor is encoded in frequency modulation of the oscillator's output waveform. 
By demodulating the output waveform, the original time-domain magnetic field signal $B(t)$ may be recovered. We describe that process here. The oscillator's instantaneous phase $\phi(t)$ is governed by Eqn.~\ref{eqn:oscillatorphasevsBfield},
\begin{equation*}
\phi(t) = \int_0^t \gamma \left[B_0+ B_\text{sen}(\tau)\right] d\tau,
\end{equation*}
where we assume operation with a sufficiently large bias field to saturate the YIG's magnetization. Differentiating the oscillator's instantaneous phase yields
\begin{equation}
\frac{d\phi(t)}{dt} = \gamma \left[B_0+B_\text{sen}(t)\right].
\end{equation}
The time domain magnetic field waveform $B_\text{sen}(t)$ is then determined by calculating 
\begin{equation}\label{eqn:recovermagneticfield}
B_\text{sen}(t) = \frac{1}{\gamma}\frac{d\phi(t)}{dt}-B_0.
\end{equation}
As a practical matter, we note the demodulation process is facilitated by applying the Hilbert transform to the (real-valued) voltage waveform of the oscillator, producing a complex signal that allows the instantaneous phase $\phi (t)$ to be determined in a simple manner. The phase is then ``unwrapped'' if necessary so that it is continuous and free from $2\pi$ jumps, and finally $\frac{d\phi(t)}{dt}$ is calculated numerically using the difference in phase between successive points in the digitized signal.

\subsection{Hilbert transform properties}\label{sec:propertieshilberttransform}

The demodulation scheme described above determines the value of the magnetic field from the instantaneous phase of the oscillator. We now detail how the Hilbert transform allows the instantaneous phase $\phi(t)$ to be determined, and in particular how $\phi(t)$ is isolated from variations in the instantaneous amplitude. Use of the Hilbert transform to achieve this objective requires two conditions be met: first, the additive phase noise $\varphi(t)$ must be small, i.e. $|\varphi(t)| \ll 1$, and second, the additive phase noise $\varphi(t)$ and additive amplitude noise $\alpha(t)$ must vary slowly compared to the intermediate frequency out of the mixer $\omega_i$ (that is, both must have negligible frequency components above $\omega_i$). Both conditions hold for the device in this work. 

The output of the mixer is digitized and may be written as a real-valued waveform,
\begin{equation}\label{eqn:generaloscillator}
v(t) = V_0[1 + \alpha(t)] \cos[\omega_i t + \varphi(t)].
\end{equation}
Given $|\varphi(t)| \ll 1$ (the first condition), trigonometric identities and the small angle approximations ($\cos \varphi(t) \approx 1$ and \mbox{$\sin \varphi(t) \approx \varphi(t)$}) allow Eqn.~\ref{eqn:generaloscillator} to be rewritten as
\begin{equation}
v(t) \approx V_0[1 + \alpha(t)] \big[\cos[\omega_i t] - \varphi(t)\sin[\omega_i t] \big].
\end{equation}
From Bedrosian's theorem, the second condition (that $\alpha(t)$ and $\varphi(t)$ vary slowly compared to $\omega_i$) allows the Hilbert transform of $v(t)$ to be calculated by transforming only the high-frequency components $\cos[\omega_i t]$ and $\sin[\omega_i t]$ \cite{schreier2010statistical}. Denoting the Hilbert transform of $v(t)$ as $\hat{v}(t)$, we have
\begin{equation}
\hat{v}(t) \approx V_0[1 + \alpha(t)] \big[\sin[\omega_i t] + \varphi(t)\cos[\omega_i t] \big].
\end{equation}
Again using small angle approximations and trigonometric identities, we obtain the resulting signal,
\begin{equation}\label{eqn:analyticsignal}
v(t) + i \hat{v}(t) \approx V_0[1 + \alpha(t)]e^{i(\omega_i t + \varphi(t))}.
\end{equation}
The instantaneous phase of the mixed-down signal $\omega_i t + \varphi(t)$ is easily determined by taking the argument of the above. As the quantity $[1 + \alpha(t)]$ is common to both the real and imaginary components, the additive amplitude noise $\alpha(t)$ is thereby isolated from the instantaneous phase. This approach should be compared to the real-valued waveform of Eqn.~\ref{eqn:generaloscillator} where there is no direct way to isolate the instantaneous phase from additive amplitude noise. 

Note that $\varphi(t)$ is real, so its double-sided power spectrum is symmetric about zero frequency. Therefore, in the frequency domain picture, it is clear we cannot gain any sensitivity by processing both the positive and negative frequency sidebands, as their information is redundant. 

\subsection{\textcolor{john}{Test field calibration}}\label{sec:testfieldcalibration}

\textcolor{john}{The test field $B_\text{sen}^\text{rms}$ is created by a single-turn coil near the sensor. The coil is connected in series with a 50 $\Omega$ resistor and driven by a function generator. The value of $B_\text{sen}^\text{rms} = 0.9$ pT used to evaluate  sensitivity is checked by increasing the function generator's voltage by $10^3\times$ and measuring the field with a commercial magnetic field probe (Beehive Electronics, 100C EMC Probe). The commercial probe measures an rms field of $1.2 \pm 0.4$ nT, with nearly all uncertainty attributed to the probe's specified calibration uncertainty. This measurement suggests the field during testing is $1.2 \pm 0.4$ pT rms, consistent with the expected value of 0.9 pT rms using our sensor's known response tied to the electron gyromagnetic ratio.}

\subsection{Crystal anisotropy \textcolor{mhs}{and frequency shifts}}\label{sec:anisotropy}

Due to the Coulomb interaction, the wavefunctions of unpaired electrons within a crystal lattice deviate from those of an isolated atom. The distorted spatial wavefunctions couple to the electron spin via the spin-orbit interaction, breaking the isotropy of the spin Hamiltonian. This anisotropy causes the crystal to magnetize more easily along certain directions, giving rise to easy and hard magnetization axes, and introduces a crystal-orientation-dependent term into the FMR frequency. Although calculating the value of this term for the general case of an arbitrary-direction magnetic field is somewhat involved~\cite{tokheim1971optimum}, the calculation simplifies for external magnetic fields confined to lie in the $\{110\}$ plane. Under these conditions, the uniform mode resonant frequency differs from $\omega_y = \gamma B$ and is instead given to good approximation by~\cite{yager1950ferromagnetic,masters1960instability,clark1963temperature}
\begin{equation}\label{eqn:anisotropy}
\omega_y \approx \gamma \left[B +\frac{K_1}{\mu_0 M_s}\left(2+\frac{15}{2}\sin^4\theta -10\sin^2\theta\right)\right],
\end{equation}
where $\theta$ is the angle in the $\{110\}$ plane between the \mbox{$<\!100\!>$} crystallographic axis and the externally applied magnetic field, and $\frac{K_1}{\mu_0 M_s}\approx -4.2$ mT for YIG~\cite{craik1975magnetic}. Equation~\ref{eqn:anisotropy} suggests that aligning the \mbox{$< \! 111 \! >$} axis parallel to $\mathbf{B}$ will result in a resonant frequency higher than $\gamma B$ by $\approx 2\pi \times 160$ MHz, while alignment of the hard axis \mbox{$<\!100\!>$} would result in a resonance lower by $\approx 2\pi\times 240$ MHz. The dependence of the FMR frequency on the value of $K_1/M_s$ is approximately removed for \mbox{$\theta = \arcsin\left[\sqrt{(10-2\sqrt{10})/15}\right] \approx 29.7^\circ$}; this angular alignment, known as zero temperature compensation (ZTC), is employed in this work. 


As the anisotropic contribution to the FMR frequency in Eqn.~\ref{eqn:anisotropy} is additive, anisotropy-induced frequency shifts are not expected to alter the device response to AC magnetic fields to first order in $B_\text{sen}/B_0$. Higher-order anisotropic effects also exist beyond those included in Eqn.~\ref{eqn:anisotropy}, but these effects are considered to be negligible for YIG~\cite{tokheim1971optimum}.


\textcolor{john}{The ZTC alignment described above can mitigate temperature-induced frequency shifts of the FMR; for example, the YIG sphere's temperature might vary as the power applied to the YIG sphere fluctuates. However, fluctuations in applied power could shift the YIG frequency in other ways, for example by reducing $M_z$ as the magnetization is tipped away from the z-axis. Here, Kittel's formula in the main text illustrates a key advantage for a sphere over other geometries such as a rod or plane~\cite{anderson1955instability}; only for a sphere is the FMR frequency independent of the magnetization. Thus a sphere protects against frequency shifts due to changes in the magnetization along $\hat{z}$ from fluctuations in applied power.}

\subsection{Fundamental limits for a spin magnetometer}

The spin-projection-limited magnetic sensitivity $\eta_\text{spl}$ for a spin-based DC magnetometer is ~\cite{budker2007optical,budker2020sensing}
\begin{equation}\label{eqn:spl}
\eta_\text{spl} \approx \frac{\hbar}{g_e \mu_B}\frac{1}{\sqrt{N T_2^*}},
\end{equation}
where $N$ is the number of total spins and $T_2^*$ is the free induction decay time (i.e. dephasing time). Importantly, Eqn.~\ref{eqn:spl} assumes the $N$ spins are independent. In YIG, there are $4.22\times 10^{21}$ unit formula of Y$_3$Fe$_5$O$_{12}$ per cm$^3$, with each unit formula contributing 5 unpaired electrons. For a 1 mm diameter YIG sphere at room temperature~\cite{anderson1964molecular}, the number of unpaired spins is $N=8\times 10^{18}$. For a FWHM unloaded linewidth of $2\pi\times 560$ kHz ($T_2^* \approx 570$ ns), we have $\eta_\text{spl}= 2.7$ aT$\sqrt{\text{s}}$. 

The relevance of this expression as a measure of the fundamental limits of a ferrimagnetic magnetometer remains unclear, as the strong coupling of nearby spins in a ferrimagnet violates the assumption of independent spins. Indeed, the extremely low spin-projection limit calculated for YIG highlights that the limits of this type of magnetometer likely must be understood quite differently than those of its paramagnetic counterparts. While the coupling present in ferrimagnets may allow entanglement-enhanced sensing schemes which surpass the limit imposed by Eqn.~\ref{eqn:spl}~\cite{kimball2016precessing,huelga1997improvement}, other expressions may emerge with further study that produce less optimistic fundamental limits. 



\textcolor{mhs}{We now examine one such additional limit on device performance that may prove to be fundamental.} \textcolor{john}{Thermal variations in the YIG sphere's magnetization are expected to translate to variations in the sphere's magnetic field. If large enough, such variations could limit the sensitivity of the device. Using the fluctuation-dissipation theorem, the expected thermal magnetic noise can be \textcolor{mhs}{estimated} as \cite{brown1963thermal,smith2002fluctuation,smith2001modeling,smith2006comment,smith2001whitenoise,safonov2005fluctuation}
\begin{equation}\label{eqn:thermalnoise}
    \eta_\text{the} \approx \sqrt{\frac{k_B T}{\gamma M_s V_y Q_0}},
\end{equation}
where $V_y$ is the volume of the YIG sphere. Evaluation at room temperature with $M_s= 142000$ A/m for a 1 mm diameter YIG sphere with $Q_0=5$ GHz $/$ 560 kHz = 8900 gives $\eta_\text{the} \approx 190$ aT$\sqrt{\text{s}}$. \textcolor{mhs}{Under these conditions, this} noise is approximately 70 times higher than the spin projection noise calculated above. \textcolor{mhs}{We note that} several papers employ a similar formula \cite{vetoshko2003epitaxial,balynsky2017magnetometer} to Eqn.~\ref{eqn:thermalnoise} but with different prefactors of order unity, \textcolor{mhs}{so that the expression above is best understood as a rough estimate of the thermal noise limit.} By convention the theoretical limits for $\eta_\text{spl}$ and $\eta_\text{the}$ are referenced to a 1 second measurement rather than a 1 Hz bandwidth \textcolor{mhs}{(as indicated by the units)}; see Ref.~\cite{fescenko2020diamond}.}

\textcolor{mhs}{Finally, regardless of the fundamental limits, more restrictive practical limits may emerge. For example,} the coupling between spins may produce limits on the power that can be applied to probe the resonance, as this coupling gives rise to degenerate spin-wave modes coupled to the uniform precession mode~\cite{suhl1956nonlinear}.

\subsection{Errors introduced by the finite bias field magnitude}\label{sec:finitebiaserrors}

Nominally, the sensor measures the projection of the external magnetic field $\mathbf{B}_\text{sen}$ along the direction of the bias magnetic field $\mathbf{B}_0$ created by the permanent magnet pair, so that the device operates as a vector magnetometer. However, slight errors are introduced by components of $\mathbf{B}_\text{sen}$ orthogonal to the bias magnetic field $\mathbf{B_0}$. We analyze the origin and magnitude of this error here.

The total field seen by the sensor arises both from the permanent magnet which is defined to produce field $B_0$ along the $\hat{z}$ direction (i.e. $\mathbf{B}_0 \equiv B_0 \hat{z}$) and the field to be sensed, $\mathbf{B}_\text{sen}$. The MW magnetic field applied at the FMR frequency is assumed to oscillate rapidly compared to the sensing bandwidth, allowing its contribution to be neglected. With this approximation, the total field seen by the sensor is
\begin{equation}
\mathbf{B} = \mathbf{B}_0+ \mathbf{B}_\text{sen}.
\end{equation}
Neglecting the perturbative effects of crystal anisotropy (see \textcolor{mhsnew}{SM~Sec.~L}), the FMR frequency depends only on $|\mathbf{B}|$. In the limit where $\mathbf{B}_0 \gg \mathbf{B}_\text{sen}$, the device responds linearly to the component of $\mathbf{B}_\text{sen}$ parallel to $\mathbf{B}_0$ and responds weakly to the component of $\mathbf{B}_\text{sen}$ perpendicular to $\mathbf{B}_0$, as will be shown. The external field may be decomposed as $\mathbf{B}_\text{sen} = \mathbf{B}_\text{sen}^\parallel + \mathbf{B}_\text{sen}^\perp$, where $B_\text{sen}^\parallel$ and $B_\text{sen}^\perp$ are the external field components parallel and perpendicular to $\mathbf{B_0}$, respectively. The scalar field is then
\begin{align}
B & = \sqrt{\mathbf{B} \cdot \mathbf{B}} \\
&= \sqrt{\left(B_0+B_\text{sens}^\parallel \right)^2 + \left( B_\text{sens}^\perp \right)^2} \\
& = B_0 \sqrt{\left( 1 + \frac{B_\text{sens}^\parallel}{B_0}\right)^2 + \left( \frac{B_\text{sens}^\perp}{B_0} \right)^2} \\
& = B_0 \sqrt{ 1 + 2 \frac{B_\text{sens}^\parallel}{B_0} + \left(\frac{B_\text{sens}^\parallel}{B_0}\right)^2+\left(\frac{B_\text{sens}^\perp}{B_0} \right)^2}.
\end{align}
Taylor expanding the final expression above with \mbox{$B_\text{sen}\ll B_0$}, we have 
\begin{equation}
B \approx B_0 + B_\text{sen}^\parallel + \frac{\left(B_\text{sen}^\perp\right)^2}{2 B_0}.
\end{equation}
The third term in the above expansion provides an estimate of the error,
\begin{equation}
\text{error} \approx \frac{\left(B_\text{sen}^\perp\right)^2}{2 B_0},
\end{equation}
 The maximum error occurs when $\mathbf{B}_\text{sen}$ is oriented perpendicular to the z axis. The error is eliminated when $\mathbf{B}_\text{sen}$ is parallel to the z-axis. For a $0.05$~mT external field and $|\mathbf{B}_0| = 0.178$~T, the maximal error is 7~nT.

These errors may be suppressed (by $\sim 10^3$ or more) by constructing a full vector magnetometer out of three sensors. In this configuration, the measured values from each of the three sensors would be combined to refine the reconstructed magnetic field vector.

\end{document}